\documentclass[10pt, twocolumn, comsoc]{IEEEtran}

\usepackage{graphicx,epsfig}

\usepackage[noadjust]{cite}
\usepackage{mcite}
\usepackage{amsfonts,helvet}
\usepackage{fancyhdr}
\usepackage{threeparttable}
\usepackage{epsf,epsfig}
\usepackage{amsthm}
\usepackage{amsmath}
\usepackage{siunitx}
\usepackage{amssymb}
\usepackage{stfloats}

\usepackage[colorlinks=true, linkcolor=blue]{hyperref}

\usepackage{dsfont}
\usepackage{subfigure}
\usepackage{color}
\usepackage{enumerate}
\usepackage{gensymb}
\usepackage{cancel}
\usepackage{lipsum}
\usepackage{mathtools}
\usepackage{cuted}
\usepackage{bbm}
\usepackage[linesnumbered,ruled]{algorithm2e}

\newtheorem{corollary}{Corollary}

\newtheorem{proposition}{Proposition}

\usepackage{eucal}
\usepackage{booktabs}
\begin{document}

\title{MIMO Integrated Sensing and Communications: \\A Fisher-based Local Mutual Information Approach}

    \author{Hyeonuk~Kim, Jeongwoo~Park, Nuria~González-Prelcic, Chan-Byoung~Chae and Jeonghun~Park

\thanks{This work was supported by the Institute of Information \& Communications Technology Planning \& Evaluation (IITP) grants funded by the Korea government (MSIT) under Grant IITP-2025-RS-2024-00428780 (6G Cloud Research and Education Open Hub), Grant RS-2024-00397216 (Development of the Upper-mid Band Extreme massive MIMO (E-MIMO)), and Grant IITP-2024-RS-2024-00434743 (YKCS Open RAN Global Collaboration Center).}
\thanks{H. Kim, J. Park, and J. Park are with the School of Electrical and Electronic Engineering, Yonsei University, South Korea (e-mail: garksi11@yonsei.ac.kr; simyffff@yonsei.ac.kr; jhpark@yonsei.ac.kr).}
\thanks{N. Gonz\'alez-Prelcic is with the Department of Electrical and Computer Engineering, University of California San Diego, La Jolla, CA 92093 USA (e-mail: ngprelcic@ucsd.edu).}
% TODO(학회판): \thanks{Part of this work was presented at [conference], [year] \cite{[conf_key]}.}
\thanks{C.-B. Chae is with the School of Integrated Technology, Yonsei University, South Korea (e-mail: cbchae@yonsei.ac.kr).}
}

\maketitle \setcounter{page}{1} 
\begin{abstract}
Integrated sensing and communications (ISAC) beamforming requires a sensing utility that is compatible with the communication rate and aligned with per-target angle estimation. Fisher information (FI)/Cram\'er--Rao bound (CRB) criteria depend on the angular units and generally involve coupled multi-target matrices, whereas sensing mutual information (SMI) is typically defined on aggregate response variables. This paper proposes Fisher-based local mutual information (F-LMI), a prior-aware bit-domain utility for multi-user multiple-input multiple-output (MU-MIMO) ISAC. F-LMI linearizes each target-specific range-compressed statistic around the angular prior mean and yields a closed-form local information gain. In the single-target case, it admits a Bayesian CRB (BCRB) log-ratio interpretation and is a locally consistent surrogate for the nonlinear parameter MI. For multiple targets, the prior-induced variations of the non-intended targets are marginalized into a beamformer-dependent interference-plus-noise covariance in each branch. We derive a fractional-programming algorithm with linear-system updates and scalar bisection. Monte Carlo simulations with nonlinear target-wise posterior-mean receivers show that F-LMI reduces the root-mean-square error (RMSE) by 5.0--14.7\% relative to a target-indexed Bayesian-bound baseline at two matched rates in a closely spaced layout. Under a common basis-selection protocol, it also attains lower branch RMSE than aggregate and joint criteria.
\end{abstract}


\section{Introduction}

% Integrated sensing and communications (ISAC) has emerged as a key enabling technology for future wireless systems, where communication and sensing share spectrum, hardware, and also spatial degrees of freedom \cite{liu_tcomm2020_road,LiuJSAC2022_ISAC}. 

\IEEEPARstart{I}{ntegrated}  sensing and communications (ISAC) has emerged as a key enabling technology for future wireless systems, where communication and sensing share spectrum, hardware, and spatial degrees of freedom \cite{liu_tcomm2020_road,LiuJSAC2022_ISAC}. 
Recent dual-function designs co-optimize the waveform, beamformer, and radio-frequency (RF) chain \cite{LiuTWC18,LiuTSP18_DFRC_OptimalWaveform}. 
In multi-user multiple-input multiple-output (MU-MIMO) ISAC, spatial beamforming is jointly exploited for user multiplexing and target probing, enabling improved spectral efficiency, hardware reuse, and design flexibility \cite{LiuTWC18,liu_tsp2020_joint,LiuTSP18_DFRC_OptimalWaveform,HuaTVT2023_TypeII}. In such systems, the key design challenge lies in the inherent communication--sensing 
performance tradeoff, which has become a central problem in ISAC beamforming 
design \cite{xiong_tit2023_gaussian,HuaTWC24_CRBRate,RenTWC24_MultiTargetCRBRate}. 
Addressing this tradeoff, in turn, requires a careful choice of the performance measures for both communication and sensing.

Unlike the communication side, where the achievable sum-rate provides a standard bit-domain performance measure, the sensing side admits several task-dependent criteria, including beampattern matching, signal-to-clutter-plus-noise ratio (SCNR), Fisher information (FI)/Cram\'er--Rao bound (CRB)-type measures, and sensing mutual information (SMI) \cite{4276989,LiuTSP22_CRB_JRC,HuaTWC24_CRBRate,RenTWC24_MultiTargetCRBRate,10411942,10255745,10977963,10584278}. 
% In particular, estimation-theoretic criteria such as FI, CRB, and their Bayesian counterparts are attractive because they are directly tied to parameter-estimation accuracy \cite{VanTrees2002,LiuTSP22_CRB_JRC,HuaTWC24_CRBRate,RenTWC24_MultiTargetCRBRate,10584278,6541985}. 
Among these, estimation-theoretic criteria such as FI, CRB, and their Bayesian counterparts are attractive because they characterize parameter-estimation error limits under the
corresponding regularity conditions  \cite{VanTrees2002,LiuTSP22_CRB_JRC,HuaTWC24_CRBRate,RenTWC24_MultiTargetCRBRate,10584278,6541985}. 
% However, they are expressed in an inverse-error domain rather than in bits, so a weighted sum with communication rate must simultaneously encode design preference and reconcile incompatible units. 
Raw FI and CRB values depend on the angular units, so their weights in a joint communication--sensing objective must be rescaled when those units change. Multi-target formulations also require scalarizing an information matrix for beamforming design.
% Moreover, in multi-target settings, FI/CRB formulations typically involve coupled information matrices whose dimension and interpretation become increasingly cumbersome as the number of sensing parameters grows. 

% Information-theoretic sensing metrics address part of this difficulty by measuring sensing utility in bits. 

Information-gain metrics quantify uncertainty reduction independently of the angular units and, after normalization by the sensing block length, are expressed in bits per channel use.
This idea originates in mutual information (MI)-based radar waveform design \cite{YangBlumTAES07_MI_MMSE,6541985}, where the waveform is chosen to maximize the MI between the observation and the target response. Recently, it has been adopted in ISAC joint waveform and beamforming optimization \cite{10255745,10411942,10554642,10977963}.
% MI-based radar waveform design has been studied extensively \cite{YangBlumTAES07_MI_MMSE,TangLiTSP19_MI_Radar,5424056,6541985}, and recent ISAC works have adopted SMI-type objectives for joint waveform or beamforming optimization \cite{10255745,10411942,10554642,10977963}. 
To keep the MI expression tractable, in this approach, the target response matrix is typically modeled as a random matrix with a known covariance, and the resulting SMI measures the information that the observation carries about that random response.
In line-of-sight (LoS)- or specular-dominant point-target sensing, however, the response is not an unstructured random matrix but a deterministic function of the target angle, e.g., $\mathbf{A}(\theta)=\alpha\mathbf{b}(\theta)\mathbf{a}^{\sf H}(\theta)$, so that all angular information is carried by the scalar $\theta$ \cite{4350230,LiTSP08_CRB_MIMORadar,10584278}.
Since the observation depends on the angle only through the response, response-level MI is not intrinsically misaligned with angle estimation; what the Gaussian random-response model does not provide is a quantification of its error with respect to the angle MI, and the resulting utility is defined on the aggregate transmit covariance rather than on the statistic formed for each target.

Motivated by the relation between MI and estimation in Gaussian models
\cite{1412024,1564431}, we develop Fisher-based local mutual
information (F-LMI), a prior-aware sensing utility defined on
target-specific range-compressed statistics.
Linearizing each range-compressed response around the angular prior mean
yields a local Gaussian information measure governed by the
array-response Jacobian.
F-LMI builds on established connections between sensing information and
estimation. Estimation-rate formulations relate radar parameter
uncertainty to information measures
\cite{ChiriyathTSP16_EstRate,ChiriyathTCCN17_Convergence}, with a
rate--distortion perspective developed in \cite{DongTVT23_EstRate}. For
linear Gaussian models, the logarithmic reduction of posterior variance
is the scalar Bayesian D-optimal information gain
\cite{ChalonerVerdinelli95,Alexanderian22_Dopt}. Prior-aware angular design has been studied using a posterior CRB (PCRB) under an unknown deterministic reflection coefficient \cite{10584278}. Related multi-target studies consider sum-PCRB covariance optimization \cite{YaoZhang_MultiTarget} and periodic-PCRB multi-user ISAC beamforming with dedicated sensing beams \cite{YaoZhang_HowManyBeams}. These multi-target studies use independent zero-mean random reflectivities. Weighted communication--sensing MI design is studied in \cite{10554642}.

Our known-reflectivity model generally retains angular coupling
in assigned range-compressed outputs. F-LMI combines intended
sensitivity with a beamformer-dependent covariance of neighboring
angular variations. Omitting nuisance terms yields noise-only
Bayesian D-optimal utilities; the target-indexed baseline uses
prior-normalized A-optimal aggregation. We establish fixed-beamformer
single-target consistency with nonlinear MI in the stated local
regime and derive a fractional-programming (FP) joint beamforming algorithm
\cite{ShenTWC18,ShenToN19_MatrixFP,ChenTWC25_FastFP_FI}.
 
% In F-LMI, the key idea is to leave the structured response deterministic in $\theta$, place the prior on $\theta$ itself, then linearize the target-specific matched-filtered response around the angular prior mean. 

% The key idea behind F-LMI is to keep the array-manifold structure of the response intact—modeling it as a deterministic function of $\theta$ instead of an unstructured random matrix—while placing the prior on $\theta$ itself, and then to linearize the target-specific matched-filtered response around the angular prior mean. 

The main contributions are summarized as follows:
\begin{itemize}
    \item We develop a receiver-oriented multi-target F-LMI utility for
    a receiver that estimates each target from its assigned
    range-compressed output. The local variations of the
    non-intended target angles are marginalized into a
    beamformer-dependent interference-plus-noise covariance; the
    construction retains a Bayesian local-information interpretation and
    reduces to per-target D-optimal utilities with a noise-only
    covariance when these nuisance terms are omitted.

    \item For a fixed sensing beamformer, we establish an explicit local
    error bound relating F-LMI to the nonlinear range-compressed MI under
    a non-degenerate information regime, and derive the corresponding
    local BCRB log-ratio interpretation.

    \item We derive an FP algorithm for the proposed beamformer-dependent
    branch utilities, with closed-form auxiliary updates, linear-system
    beamformer updates, and scalar bisection. We evaluate the resulting
    designs through rate--root-mean-square error (RMSE) comparisons in
    close- and well-separated layouts, a target-separation sweep, and a
    common basis-selection experiment, with a common receiver across
    methods within each experiment.
\end{itemize}

\section{System Model}
\label{sec:system_model}
% Cramér–Rao lower bound (CRLB) and mutual information (MI) are often studied separately, which can obscure their connection. We therefore state the system model in a way that makes the CRLB–MI link explicit.

Consider a downlink MU-MIMO ISAC system in which an $N_T$-antenna base station (BS) serves $K$ single-antenna users while sensing $M$ distinct point targets.
Over a block of length $L$, the BS transmits
\begin{align}
    \mathbf{X} = \mathbf{W}_c \mathbf{S}_c + \mathbf{W}_s \mathbf{S}_s,
    \label{eq:signal}
\end{align}
where $\mathbf{W}_c \in \mathbb{C}^{N_T \times K}$ and $\mathbf{W}_s \in \mathbb{C}^{N_T \times M}$ denote the communication and sensing precoders, respectively. 
% The symbol matrices $\mathbf{S}_c \in \mathbb{C}^{K \times L}$ and $\mathbf{S}_s \in \mathbb{C}^{M \times L}$ contain i.i.d. $\mathcal{CN}(0,1)$ entries and are mutually independent. 
The entries of the communication and sensing symbol matrices $\mathbf{S}_c \in \mathbb{C}^{K \times L}$ and $\mathbf{S}_s \in \mathbb{C}^{M \times L}$ are independently drawn from $\mathcal{CN}(0,1)$, and the two matrices are mutually independent. 
% {\color{red}{XXX}}
Since the communication and sensing streams are mutually independent with
unit variance, the transmit covariance is
\begin{align}
\mathbf{R}_x 
\triangleq 
\frac{1}{L}\mathbb{E}[\mathbf{X}\mathbf{X}^{\sf H}]
=
\mathbf{W}_c\mathbf{W}_c^{\sf H}
+
\mathbf{W}_s\mathbf{W}_s^{\sf H}.
\end{align}
For finite but sufficiently large $L$, the sample covariance satisfies
\begin{align}
\frac{1}{L}\mathbf{X}\mathbf{X}^{\sf H}
\approx
\mathbf{R}_x
\end{align}
under the assumed independent Gaussian signaling.
The considered system is subject to the average total transmit-power constraint, given by
\begin{align}
\mathrm{tr}(\mathbf{R}_x)=\|\mathbf{W}_c\|_F^2+\|\mathbf{W}_s\|_F^2 \le P_{\mathrm{tx}}.
\end{align}
This transmission model is standard in the ISAC literature \cite{liu_tsp2020_joint,xiong_tit2023_gaussian,liu_tcomm2020_road,LiuJSAC2022_ISAC,LiuTWC18}.

\subsection{Communication Model}
The received signal matrix at the users, denoted as ${\bf{Y}}_c \in \mathbb{C}^{K \times L}$, is given by
\begin{align}
    \mathbf{Y}_c = \mathbf{H}\mathbf{X} + \mathbf{Z}_c = \mathbf{H}(\mathbf{W}_c \mathbf{S}_c + \mathbf{W}_s \mathbf{S}_s) + \mathbf{Z}_c,
\end{align}
where $\mathbf{H} \in \mathbb{C}^{K \times N_T}$ denotes the downlink channel matrix between the BS and the $K$ users and ${\bf{Z}}_c \in \mathbb{C}^{K \times L}$ is the additive Gaussian noise whose entries are independent and identically distributed complex Gaussian random variables with zero mean and variance $\sigma_c^2$. 

Let $\mathbf{h}_k^{\sf H}$ denote the $k$-th row of $\mathbf{H}$, and
let $\mathbf{s}_{c,k}\triangleq\mathbf{S}_c(k,:)$ denote the data
symbol sequence intended for user $k$. The received sequence at user
$k$ is
\begin{align}
\mathbf{y}_{c,k} 
&= \mathbf{h}_k^{\sf H}\mathbf{X} + \mathbf{z}_{c,k} \nonumber \\
&= \mathbf{h}_k^{\sf H}\mathbf{w}_{c,k}\mathbf{s}_{c,k} + \sum_{j\neq k} \mathbf{h}_k^{\sf H}\mathbf{w}_{c,j}\mathbf{s}_{c,j} + \mathbf{h}_k^{\sf H}\mathbf{W}_s\mathbf{S}_s + \mathbf{z}_{c,k},
\end{align}
where $\mathbf{z}_{c,k}$ is the $k$-th row of $\mathbf{Z}_c$.

We consider the case where the dedicated sensing waveform matrix
$\mathbf{S}_s$ is not available at the communication users. Hence, its
contribution cannot be canceled at the user receivers and is treated as
additional Gaussian interference. Under Gaussian signaling and
single-user decoding with interference treated as noise, the signal-to-interference-plus-noise ratio (SINR) for user $k$ is given by
\begin{align}
\gamma_{c,k} \triangleq \frac{|\mathbf{h}_k^{\sf H}\mathbf{w}_{c,k}|^2}{\sum_{j\neq k} |\mathbf{h}_k^{\sf H}\mathbf{w}_{c,j}|^2 + \|\mathbf{h}_k^{\sf H}\mathbf{W}_s\|_2^2 + \sigma_c^2}.
\label{eq:SINR_c}
\end{align}
Accordingly, an achievable rate of user $k$ is formulated as
\begin{align}
R_k 
&\triangleq \frac{1}{L} I \left( \mathbf{y}_{c,k}; \mathbf{s}_{c,k} \right) \nonumber \\
&= \log_2 \left( 1 + \gamma_{c,k} \right).
\label{eq:CMI}
\end{align}
We note that, alternatively, if the sensing symbols $\mathbf{S}_s$ are
known to the users a priori and can be canceled before decoding
\cite{LiuJSAC2022_ISAC,HuaTVT2023_TypeII}, the sensing-interference term
$\|\mathbf{h}_k^{\sf H}\mathbf{W}_s\|_2^2$ is removed from the
denominator of \eqref{eq:SINR_c}. The proposed F-LMI framework applies to both cases.
% We note that it is also feasible that each user has prior knowledge of ${\bf S}_s$, which enables the sensing interference to be canceled before decoding \cite{LiuJSAC2022_ISAC, HuaTVT2023_TypeII}. 
% We clarify that our framework is also applicable in the case where each user is capable of removing ${\bf{S}}_s$. 

% The term $\mathbf{Z}_C \in \mathbb{C}^{K \times L}$ is the additive noise matrix whose entries are i.i.d. complex Gaussian random variables with zero mean and variance $\sigma_c^2$.
% Note that in (7), the term $\mathbf{H}\mathbf{W}_s \mathbf{S}_s$ represents the interference caused by the dedicated sensing beams.

\subsection{Sensing Model}

For sensing, the BS employs a separate receive array co-located with its transmit array. Following the known-waveform sensing model \cite{10411942,10255745,xiong_tit2023_gaussian}, the receiver knows the realized communication symbols $\mathbf S_c$, sensing sequences $\mathbf S_s$, and precoders, and hence the transmitted block $\mathbf X$. The target responses are constant over the synchronized observation block. Receiving during transmission implies in-band full-duplex operation; residual self-interference is assumed to be suppressed below the noise floor \cite{BayraktarTWC26_FDISAC}. The $M$ targets lie within the same range bin and must therefore be resolved in angle.

% As our primary objective is to propose a new sensing metric, we employ a unified formulation without explicitly differentiating between the two cases. Both monostatic and bistatic systems can be accommodated by appropriately configuring the steering vectors and system parameters. 

The sensing receiver observes the reflected echoes from $M$ point targets through an $N_R$-antenna receive array. The received signal matrix $\mathbf{Y}_R \in \mathbb{C}^{N_R \times L}$ is expressed as
\begin{align} \label{eq:radar_obs_model}
    \mathbf{Y}_R = \sum_{m=1}^{M} \mathbf{A}(\theta_m)\mathbf{X} + \mathbf{Z}_R,
\end{align}
where $\mathbf Z_R$ has i.i.d. $\mathcal{CN}(0,\sigma_z^2)$ entries and is independent of the transmitted symbols and target angles. The target response matrix $\mathbf{A}(\theta_m) \in \mathbb{C}^{N_R \times N_T}$ is defined as
\begin{align} \label{eq:At_def}
    \mathbf{A}(\theta_m) \triangleq \alpha_m \mathbf{b}(\theta_m)\mathbf{a}^{\sf H}(\theta_m),
\end{align}
where $\alpha_m$ is the complex reflection coefficient of the $m$-th target.
We note that this structured point-target response model is standard in colocated MIMO radar and array-manifold-based sensing formulations \cite{4350230,LiTSP08_CRB_MIMORadar}. 
$\mathbf{a}(\theta_m) \in \mathbb{C}^{N_T}$ and $\mathbf{b}(\theta_m) \in \mathbb{C}^{N_R}$ represent the transmit and receive steering vectors corresponding to the $m$-th target located at direction $\theta_m$. 

We focus on the estimation of the target directions $\{\theta_m\}_{m=1}^M$ and treat the reflection coefficients $\{\alpha_m\}$ as deterministic and known (or pre-estimated) at the receiver. In tracking-oriented scenarios, reflection coefficients can be updated between sensing intervals and held fixed during beamforming. Under this premise, prior statistical knowledge on each $\theta_m$ is incorporated through a Gaussian model $\theta_m\sim\mathcal N(\mu_{\theta_m},\sigma_{\theta_m}^2)$, where $\mu_{\theta_m}$ and $\sigma_{\theta_m}^2$ denote the prior mean and variance, respectively. Using the known transmit sequences, the receiver forms stream-specific range-compressed outputs, each retaining echoes from all targets. Each target is estimated from its assigned output, with neighboring angular variations treated as nuisance uncertainty.

\section{Preliminaries: Conventional Sensing Metrics}
\label{sec:preliminaries}

We briefly review FI/CRB- and SMI-based sensing utilities to
motivate the proposed F-LMI.

\textbf{CRB/FI-based metrics:}
For unbiased estimators of a real-valued parameter vector
$\boldsymbol{\theta}$ under the usual regularity conditions,
the CRB gives
\begin{align}
\mathbb{E}
\left[
(\hat{\boldsymbol{\theta}}-\boldsymbol{\theta})
(\hat{\boldsymbol{\theta}}-\boldsymbol{\theta})^{\sf T}
\right]
\succeq
\mathbf F^{-1}(\boldsymbol{\theta}),
\end{align}
where $\mathbf F(\boldsymbol{\theta})$ is the Fisher information matrix.
FI/CRB-based ISAC designs are attractive because they are directly tied
to parameter-estimation accuracy
\cite{VanTrees2002,LiuTSP22_CRB_JRC,HuaTWC24_CRBRate,RenTWC24_MultiTargetCRBRate,10584278}.
For a monostatic uniform linear array (ULA) MIMO radar tracking a single far-field point
target, and with $\mathbf A(\theta)$ defined as in \eqref{eq:At_def}, the known-reflectivity angle CRB reduces to
\begin{align}
    \mathrm{CRB}(\theta)=
    \frac{\sigma_z^2}
    {2L\operatorname{tr} \left(\dot{\mathbf{A}}^{\sf H}\dot{\mathbf{A}}\mathbf{R}_x\right)},
\label{eq:mimo_crb_known}
\end{align}
where $\dot{\mathbf A}\triangleq \partial \mathbf A(\theta)/\partial \theta$ and
$\mathbf R_x$ is the transmit covariance. Hence, under known reflectivity,
minimizing the local angle CRB is equivalent to maximizing the scalar FI
term
$\operatorname{tr}(\dot{\mathbf A}^{\sf H}\dot{\mathbf A}\mathbf R_x)$.
When $\alpha$ is unknown and treated as a nuisance parameter, an
additional information-loss term appears in the CRB
\cite{LiuTSP22_CRB_JRC}.

Thus, FI/CRB criteria are estimation-aligned but depend on the
parameterization and operating point. Bayesian variants average
likelihood information over the angular prior and incorporate
prior information \cite{10584278}.

\textbf{SMI-based metrics:}
SMI provides a bit-domain sensing utility and can therefore be combined naturally with communication-rate objectives.
% \footnote{
% The Kullback--Leibler divergence (KLD) is also common in sensing,
% especially for hypothesis testing or discrimination-oriented design.
% Since this work focuses on continuous angle estimation and a
% rate-compatible sensing utility, we use an MI-based formulation rather
% than a pairwise-divergence criterion.
% }
For a fixed known transmit block $\mathbf X$, the conventional Gaussian-response model treats $\mathbf g\triangleq\operatorname{vec}(\mathbf A)$ as a block-constant proper complex Gaussian vector with covariance
$\mathbf R_{\mathbf g}$. With the large-$L$ approximation $\mathbf X\mathbf X^{\sf H}/L\approx\mathbf R_x$, the SMI is commonly written as
\begin{align}
I(\mathbf Y_R;\mathbf A\mid\mathbf X)
\approx
\log_2\det\!\Big(
\mathbf I_{N_TN_R}
+\tfrac{L}{\sigma_z^2}\,(\mathbf R_x^*\otimes\mathbf I_{N_R})\,\mathbf R_{\mathbf g}
\Big),
\label{eq:smi_form}
\end{align}
where $L$ appears inside the log-determinant because all $L$ snapshots observe the same response \cite{10411942,10255745,10554642,10977963}.

Since the response is a deterministic function of the target angles,
$I(\mathbf Y_R;\mathbf A\mid\mathbf X)=I(\mathbf Y_R;\boldsymbol\theta\mid\mathbf X)$ holds under the true induced distribution, so response-level MI is not intrinsically misaligned with angle estimation. The array manifold is nonlinear in $\boldsymbol\theta$, however, so the response distribution induced by an angular prior is non-Gaussian, and \eqref{eq:smi_form} by itself neither quantifies the error of the Gaussian surrogate relative to the angle MI nor establishes an angular BCRB interpretation. Moreover, \eqref{eq:smi_form} depends on the precoders only through $\mathbf R_x$, so it scores the aggregate transmit covariance rather than the statistic formed for each target; the per-target SMI upper bound attained under cross-correlation suppression \cite{SMIUB} is defined on $\mathbf R_x$ as well.

% Consequently, aggregate SMI need not promote target-wise separability under target-specific matched-filtered processing, especially for closely spaced targets. 

\section{Fisher-based Local Mutual Information}
\label{sec:flmi}
% Addressing the limitations on the conventional sensing metrics CRB and SMI, we propose a new sensing performance metric for ISAC systems, termed F-LMI. 
% The proposed F-LMI provides a unified information-theoretic framework for sensing that is naturally expressed in bits and thus can be directly aligned with communication objectives (e.g., spectral efficiency in bits/s/Hz after appropriate normalization). In contrast to conventional sensing mutual information (SMI), which measures the information between the observation and the (often high-dimensional) unstructured radar channel response $\mathbf{A}(\theta)$, F-LMI directly targets the parameters of interest (e.g., angles) while remaining computationally tractable.

We now introduce F-LMI. A key idea behind F-LMI is to construct a prior-aware local MI utility by linearizing the sensing response around the angular prior mean. 
This yields a closed-form bit-domain metric whose gain is governed by the
local array-response sensitivity and which admits a local Bayesian
FI/BCRB interpretation. Since the unknown angles are real-valued whereas
the radar observations are complex-valued, we first use an equivalent
real-valued representation.
% and then derive F-LMI.

% Our key idea is 
%For 

\subsection{Real-Valued Representation for Improper Complex Gaussian}
To capture the second-order statistics relevant to estimating real-valued parameters from complex-valued observations, we employ the widely-linear (augmented) viewpoint \cite{Schreier2010} via a real-valued isomorphism.
Let $\mathbf{y}_R \triangleq \mathrm{vec}(\mathbf{Y}_R) \in \mathbb{C}^{N_R L}$ denote the vectorized radar observation. We define the real composite vector $\tilde{\mathbf{y}}_R \in \mathbb{R}^{2N_R L}$ by stacking the real and imaginary parts:
\begin{align}
    \tilde{\mathbf{y}}_R \triangleq 
    \begin{bmatrix} \Re\{\mathbf{y}_R\}^{\sf T}, & \Im\{\mathbf{y}_R\}^{\sf T} \end{bmatrix}^{\sf T}.
    \label{eq:real_vectorization}
\end{align}
Since the mapping $\mathbf{y}_R \mapsto \tilde{\mathbf{y}}_R$ is bijective, MI is preserved: $I(\mathbf{y}_R;\boldsymbol{\theta}) = I(\tilde{\mathbf{y}}_R;\boldsymbol{\theta})$.

Similarly, for proper complex Gaussian noise $\mathbf{z} \sim \mathcal{CN}(\mathbf{0}, \mathbf{C}_z)$, the equivalent real noise $\tilde{\mathbf{z}}$ follows $\mathcal{N}(\mathbf{0}, \mathbf{C}_{\tilde{z}})$ with covariance:
\begin{align}
    \mathbf{C}_{\tilde{z}} = \frac{1}{2}
    \begin{bmatrix}
        \Re\{\mathbf{C}_z\} & -\Im\{\mathbf{C}_z\} \\
        \Im\{\mathbf{C}_z\} & \Re\{\mathbf{C}_z\}
    \end{bmatrix}.
\end{align}
For spatially white noise $\mathbf{C}_z = \sigma_z^2 \mathbf{I}_{LN_R}$, this simplifies to $\mathbf{C}_{\tilde{z}} = \frac{\sigma_z^2}{2}\mathbf{I}_{2N_R L}$. This formulation allows us to leverage standard real-Gaussian identities in the subsequent derivation. For later use, we define for any \(\mathbf x\in\mathbb C^N\)
and \(\mathbf A\in\mathbb C^{P\times N}\)
\begin{align}
\mathcal R(\mathbf x)
\triangleq
\begin{bmatrix}
\Re\{\mathbf x\}\\
\Im\{\mathbf x\}
\end{bmatrix},\qquad
\mathcal T(\mathbf A)
\triangleq
\begin{bmatrix}
\Re\{\mathbf A\} & -\Im\{\mathbf A\}\\
\Im\{\mathbf A\} & \Re\{\mathbf A\}
\end{bmatrix}.
\end{align}
Then
\begin{align}
\mathcal R(\mathbf A\mathbf x)
=
\mathcal T(\mathbf A)\mathcal R(\mathbf x),
\qquad
\|\mathbf x\|_2^2
=
\|\mathcal R(\mathbf x)\|_2^2.
\end{align}
For Hermitian \(\boldsymbol\Phi\succeq\mathbf 0\),
\begin{align}
\mathbf x^{\sf H}\boldsymbol\Phi\mathbf x
=
\mathcal R(\mathbf x)^{\sf T}
\mathcal T(\boldsymbol\Phi)
\mathcal R(\mathbf x).
\end{align}
% \subsection{Performance Metrics}
% Performance metrics for ISAC sensing can be broadly categorized into two types: implicit metrics and explicit metrics.

% Implicit metrics, such as beampattern mismatch error, aim to shape the transmit signal to mimic a desired radar beampattern.
% While intuitive and simple to implement, these metrics serve only as indirect proxies for sensing performance.
% They do not directly quantify the estimation accuracy, often leading to suboptimal performance in complex multi-user, multi-target scenarios where interference is a dominant factor.

% Consequently, modern ISAC design prioritizes explicit metrics that directly characterize the fundamental limits of estimation.
% These explicit metrics are primarily classified into CRLB-type and SMI-type metrics.
% However, as we discuss below, both classes exhibit fundamental limitations for modern multi-user ISAC systems, particularly when estimating real-valued parameters from complex-valued observations.

% \subsection{CRLB-Type Sensing Metrics}

% \subsection{SMI-Type Sensing Metrics}

\subsection{Single-Target Scenario and Intuition}
\label{sec:flmi_single}
To explain the core idea without inter-target coupling, we first consider the single-target case ($M=1$); the multi-target setting is incorporated later.
% However, the actual beamforming design and performance evaluation will focus on the multi-target setting. 
Dropping the target index for simplicity, the observation matrix $\mathbf Y_R\in\mathbb C^{N_R\times L}$ is given by
\begin{align}
\mathbf Y_R=\mathbf A(\theta)\mathbf X+\mathbf Z_R.
\label{eq:single_target_model}
\end{align}
We form a range-compressed statistic using the known transmit sequences \cite{LiTSP08_CRB_MIMORadar}. With $\mathbf S\triangleq[\mathbf S_c^{\sf T},\mathbf s_s^{\sf T}]^{\sf T} \in\mathbb C^{(K+1)\times L}$ of full row rank ($L\ge K+1$) and $\mathbf X=[\mathbf W_c,\mathbf w_s]\mathbf S$, the dedicated sensing-stream output is
\begin{align}
\mathbf y_{\mathrm{rc}}
&\triangleq
\mathbf Y_R\mathbf S^{\sf H}
(\mathbf S\mathbf S^{\sf H})^{-1}\mathbf e_{K+1}
\nonumber\\
&=\mathbf A(\theta)\mathbf w_s+\bar{\mathbf z}_R,
\label{eq:single_target_model_mf}
\end{align}
where $\mathbf e_{K+1}$ is the last standard basis vector in $\mathbb R^{K+1}$. Conditional on $\mathbf S$, $\bar{\mathbf z}_R\sim \mathcal{CN}(\mathbf0,\sigma_{\mathrm{rc}}^2\mathbf I_{N_R})$ with $\sigma_{\mathrm{rc}}^2\triangleq \sigma_z^2[(\mathbf S\mathbf S^{\sf H})^{-1}]_{K+1,K+1}$. Under the stated block model, the retained output contains no communication-stream leakage. Its noise variance reduces to $\sigma_z^2/L$ when $\mathbf S\mathbf S^{\sf H}=L\mathbf I_{K+1}$. Conditioning on the known $\mathbf S$ is implicit in the subsequent sensing analysis unless shown explicitly. We define the sensing utility as the MI between this statistic and the target angle:
\begin{align}
I(\mathbf y_{\mathrm{rc}};\theta\mid\mathbf w_s)
&\triangleq h(\mathbf y_{\mathrm{rc}}\mid\mathbf w_s)
\nonumber\\
&\quad-h(\mathbf y_{\mathrm{rc}}\mid\theta,\mathbf w_s).
\label{eq:mi_def}
\end{align}

Since $\mathbf y_{\mathrm{rc}}$ is a deterministic function of $\mathbf Y_R$ for known $\mathbf S$, the data processing inequality gives $I(\mathbf y_{\mathrm{rc}};\theta\mid\mathbf X,\mathbf S) \le I(\mathbf Y_R;\theta\mid\mathbf X,\mathbf S)$.
The gap can be nonzero because the discarded communication-stream outputs may carry additional angular information. The MI in \eqref{eq:mi_def} does not generally admit a closed-form expression: $\mathbf A(\theta) =\alpha\mathbf b(\theta)\mathbf a^{\sf H}(\theta)$ depends nonlinearly on $\theta$, and evaluating the MI generally requires integration over the angular prior. We therefore develop a local Gaussian surrogate below.

Specifically, we apply a first-order Taylor expansion around the prior mean $\mu_\theta$ to \eqref{eq:single_target_model_mf}, which defines the local effective observation
\begin{align}
\mathbf{y}_{\text{eff}}
\triangleq \mathbf{A}(\mu_\theta)\mathbf{w}_s
+\dot{\mathbf{A}}(\mu_\theta)\mathbf{w}_s(\theta-\mu_\theta)
+ \bar{\mathbf{z}}_{R},
\label{eq:rc_local}
\end{align}
so that $\mathbf{y}_{\mathrm{rc}} \approx \mathbf{y}_{\text{eff}}$ for $\theta$ near $\mu_\theta$. Given the range-compressed model in \eqref{eq:single_target_model_mf}, all subsequent expressions are exact for the local surrogate in \eqref{eq:rc_local}; the remaining approximation is the first-order linearization.
For notational convenience, we define the local mean term as $\mathbf{m} = \mathbf{A}(\mu_\theta)\mathbf{w}_s$
and the local sensitivity (Jacobian) vector as $\mathbf{J} = \dot{\mathbf{A}}(\mu_\theta)\mathbf{w}_s$. 
% so that \eqref{eq:rc_local} can be compactly written as$\mathbf{y}_{\rm rc}\approx \mathbf{m}_{\rm rc}+\mathbf{j}_{\rm rc}(\theta-\mu_\theta)+\mathbf{z}_{\rm rc}$.
Let $\Delta\theta \triangleq \theta-\mu_\theta \sim \mathcal{N}(0,\sigma_\theta^2)$ denote the parameter deviation and define the centered range-compressed observation $\bar{\mathbf{y}}_{\text{eff}} \triangleq \mathbf{y}_{\text{eff}} -\mathbf{m}$. 
Then \eqref{eq:rc_local} is written as
\begin{align}
\bar{\mathbf{y}}_{{\text{eff}}} = \mathbf{J} \Delta\theta + \bar{\mathbf{z}}_{R}.
\label{eq:rc_centered}
\end{align}

% Although $\mathbf{z}_{\rm rc}$ is proper (circular) complex Gaussian, the marginal distribution of $\mathbf{y}'_{\rm rc}$ is generally improper
% because the random input $\Delta\theta$ is real-valued. Indeed, the pseudo-covariance of $\mathbf{y}'_{\rm rc}$ is
% \begin{align}
% \mathbf{C}_{\rm rc}
% \triangleq \mathbb{E} \left[\mathbf{y}'_{\rm rc}(\mathbf{y}'_{\rm rc})^{\sf T}\right]
% = \sigma_\theta^2\,\mathbf{j}_{\rm rc}\mathbf{j}_{\rm rc}^{\sf T}\neq \mathbf{0},
% \label{eq:rc_pseudocov}
% \end{align}
% where we used the independence of $\Delta\theta$ and $\mathbf{z}_{\rm rc}$ and the fact that $\mathbb{E}[\mathbf{z}_{\rm rc}\mathbf{z}_{\rm rc}^{\sf T}]=\mathbf{0}$ for proper noise.
% Therefore, standard mutual information formulas for proper complex Gaussian vectors are not directly applicable

The centered observation ${\bar{{\bf{y}}}}_{\text{eff}}$ is distributed as an improper complex Gaussian random vector. Using the real-valued isomorphism in \eqref{eq:real_vectorization}, we form the real-stacked counterparts 
$\tilde{\mathbf{y}}_{{\text{eff}}}$ and $\tilde{\mathbf{J}}$ of
$\bar{\mathbf{y}}_{{\text{eff}}}$ and ${\mathbf{J}}$, respectively. 
Since $\bar{\mathbf{z}}_{R}\sim\mathcal{CN}(\mathbf{0},\sigma_{\mathrm{rc}}^2\mathbf{I}_{N_R})$, its real-stacked version satisfies
$\tilde{\mathbf{z}}_{R}\sim\mathcal{N} \left(\mathbf{0},\frac{\sigma_{\mathrm{rc}}^2}{2}\mathbf{I}_{2N_R}\right)$.
Therefore, \eqref{eq:rc_centered} is equivalently represented as
the scalar-input real Gaussian channel
\begin{align}
\tilde{\mathbf{y}}_{\text{eff}}
= \tilde{\mathbf{J}}\Delta\theta + \tilde{\mathbf{z}}_{R}.
\label{eq:rc_real_channel}
\end{align}
Throughout, $\kappa\triangleq 2/\sigma_{\mathrm{rc}}^2$ denotes the noise precision of the real-stacked compressed statistic. For \eqref{eq:rc_real_channel}, the conditional MI admits the closed form as follows:
\begin{align}
I(\tilde{\mathbf{y}}_{\text{eff}};\theta|\mathbf{w}_s)
&= \frac{1}{2}\log_2 \det \left(
\mathbf{I}_{2N_R} + \sigma_\theta^2\,\tilde{\mathbf{J}}\tilde{\mathbf{J}}^{\sf T}
\left(\frac{1}{\kappa}\mathbf{I}_{2N_R}\right)^{-1}
\right) \nonumber\\
% &= \frac{1}{2}\log_2 \left(
% 1 + \sigma_\theta^2\,\tilde{\mathbf{j}}_{\rm rc}^{\sf T}
% \left(\frac{\sigma_{\rm rc}^2}{2}\mathbf{I}_{2N_R}\right)^{-1}
% \tilde{\mathbf{j}}_{\rm rc}
% \right) \nonumber\\
&= \frac{1}{2}\log_2 \left(
1+\kappa\sigma_\theta^2\,\tilde{\mathbf{J}}^{\sf T}\tilde{\mathbf{J}}
\right),
\label{eq:mi_rc_real}
\end{align}
where we use Sylvester's determinant identity. 
% for the rank-one matrix$\tilde{\mathbf{j}}_{\rm rc}\tilde{\mathbf{j}}_{\rm rc}^{\sf T}$.
Finally, using the norm-preservation property $\tilde{\mathbf{v}}^{\sf T}\tilde{\mathbf{v}}=\mathbf{v}^{\sf H}\mathbf{v}$,
we obtain 
% $\tilde{\mathbf{j}}_{\rm rc}^{\sf T}\tilde{\mathbf{j}}_{\rm rc}=\mathbf{j}_{\rm rc}^{\sf H}\mathbf{j}_{\rm rc}$, and thus
\begin{align}
I(\tilde{{\bf{y}}}_{\text{eff}};\theta|\mathbf{w}_s)
= \frac{1}{2}\log_2 \left(
1 + {\kappa\sigma_\theta^2}\,\mathbf{J}^{\sf H}\mathbf{J}
\right).
\label{eq:mi_rc_complex}
\end{align}
Substituting the definition of $\mathbf{J}$ from \eqref{eq:rc_local} into \eqref{eq:mi_rc_complex}, we obtain the final range-compressed F-LMI formulation:
% Therefore,
% $\mathbf{j}_{\rm rc}^{\sf H}\mathbf{j}_{\rm rc}
% = \mathbf{w}_s^{\sf H}\dot{\mathbf{A}}^{\sf H}(\mu_\theta)\dot{\mathbf{A}}(\mu_\theta)\mathbf{w}_s$.
% Moreover, from \eqref{eq:rc_model}, $\sigma_{\rm rc}^2=\sigma_z^2/L$, which yields the final pulse-compressed F-LMI:
\begin{align}
I(\tilde{{\bf{y}}}_{\text{eff}};\theta|\mathbf{w}_s)
&=
\frac{1}{2}\log_2 \left(
1 +\kappa\sigma_\theta^2\,
\mathbf{w}_s^{\sf H}\dot{\mathbf{A}}^{\sf H}(\mu_\theta)\dot{\mathbf{A}}(\mu_\theta)\mathbf{w}_s
\right) 
\label{eq:flmi_single_rc_final} 
\end{align}
which we denote by $I_{\mathrm{F\text{-}LMI}}(\theta)$.
For later reference, define the Fisher information of the nonlinear range-compressed model \eqref{eq:single_target_model_mf} evaluated at the prior mean as
\begin{align}
\CMcal{J}^{\mathrm{true}}(\mu_\theta)
\triangleq
\kappa
\left\|
\dot{\mathbf A}(\mu_\theta)\mathbf w_s
\right\|_2^2 ,
\label{eq:true_fi_prior_mean}
\end{align}
so that the single-target F-LMI in \eqref{eq:flmi_single_rc_final} reads
\begin{align}
I_{\mathrm{F\text{-}LMI}}(\theta)
=
\frac{1}{2}
\log_2
\left(
1+\sigma_\theta^2
\CMcal{J}^{\mathrm{true}}(\mu_\theta)
\right).
\label{eq:flmi_exact_linear_mi}
\end{align}
For fixed prior variance and the same observation model,
\eqref{eq:flmi_exact_linear_mi} is strictly increasing in
$\CMcal{J}^{\mathrm{true}}(\mu_\theta)$ and therefore preserves the
Pareto-efficient beamformers obtained with the pointwise CRB at
$\mu_\theta$ under the same feasible set. Its logarithmic form is
closely related to estimation-rate formulations and equals the Bayesian
D-optimal information gain of the local Gaussian model
\cite{ChiriyathTSP16_EstRate,Alexanderian22_Dopt}.
The expression in \eqref{eq:flmi_exact_linear_mi} is obtained from the locally linearized surrogate. In Proposition~\ref{prop:flmi_consistency}, we explicitly quantify its connection to the nonlinear range-compressed MI. Under a non-degenerate local regime in which the noise precision $\kappa$ increases while the angular uncertainty shrinks accordingly, the difference between the nonlinear range-compressed MI and F-LMI is bounded by an explicit $O(\kappa^{-1/2})$ term.
% {\color{red}{Specifically, the effective sensing information increases while the angular prior
% concentrates around its mean, so that the dimensionless local information product remains bounded and the higher-order response variation vanishes after noise whitening.}}

% {\color{red}{XXX}}

\begin{proposition}[Local MI consistency of F-LMI]
\label{prop:flmi_consistency}
Consider the nonlinear range-compressed sensing channel in \eqref{eq:single_target_model_mf} with a fixed sensing beamformer $\mathbf w_s$. Let
\begin{align}
\mathbf f(\theta)
\triangleq
\mathbf A(\theta)\mathbf w_s,
\qquad
\mathbf J(\theta)
\triangleq
\mathbf f'(\theta)
=
\dot{\mathbf A}(\theta)\mathbf w_s .
\end{align}
For each experiment $n$, condition on a known full-row-rank block
$\mathbf S_n$ and let
$\sigma_{\mathrm{rc},n}^2\triangleq\sigma_{z,n}^2
[(\mathbf S_n\mathbf S_n^{\sf H})^{-1}]_{K+1,K+1}$;
all information quantities below are conditional on $\mathbf S_n$.
Consider a sequence of experiments indexed by the noise precision
\begin{align}
\kappa_n
\triangleq
\frac{2}{\sigma_{\mathrm{rc},n}^2}
\rightarrow \infty ,
\end{align}
where the array dimensions, array geometry, prior mean, and sensing beamformer are fixed, while the prior variance $v_n\triangleq\sigma_{\theta,n}^2$ may depend on $n$. Define
\begin{align}
\CMcal{J}^{\mathrm{true}}_n(\theta)
\triangleq
\kappa_n
\|\mathbf J(\theta)\|_2^2 .
\label{eq:true_fi_def}
\end{align}
Assume that:
\begin{itemize}
    \item[(A1)] $\mathbf f:\mathbb R\to\mathbb C^{N_R}$ is twice continuously differentiable with $\sup_{\theta\in\mathbb R}\|\mathbf f''(\theta)\|_2\le C_f<\infty$. This holds automatically for the array-manifold model in \eqref{eq:At_def}, whose steering-vector entries are complex exponentials in $\sin\theta$ with bounded derivatives for fixed $N_T$, $N_R$, and fixed reflectivity.
    \item[(A2)] $\|\mathbf J(\mu_\theta)\|_2>0$, i.e., $\mathbf w_s$ provides local identifiability of $\theta$ at the prior mean.
    \item[(A3)] The dimensionless local information product satisfies
    \begin{align}
    \rho_n
    \triangleq
    v_n
    \CMcal{J}^{\mathrm{true}}_n(\mu_\theta)
    =
    \kappa_n v_n
    \|\mathbf J(\mu_\theta)\|_2^2
    =
    \Theta(1).
    \end{align}
\end{itemize}
Let $I_{\mathrm{true},n}(\theta;\mathbf y_{\mathrm{rc}}|\mathbf w_s)$ denote the MI of the nonlinear range-compressed model \eqref{eq:single_target_model_mf}, and let
\begin{align}
I_{\mathrm{F\text{-}LMI},n}(\theta)
\triangleq
\frac{1}{2}
\log_2
\left(
1+
v_n
\CMcal{J}^{\mathrm{true}}_n(\mu_\theta)
\right).
\end{align}
Then, for every $n$,
\begin{align}
\left|
I_{\mathrm{true},n}(\theta;\mathbf y_{\mathrm{rc}}|\mathbf w_s)
-
I_{\mathrm{F\text{-}LMI},n}(\theta)
\right|
\le
\frac{\sqrt{6N_R}\,C_f\,\rho_n}
{2\ln 2\,\|\mathbf J(\mu_\theta)\|_2^2}
\,
\kappa_n^{-1/2}
\label{eq:flmi_nonasymptotic_bound}
\end{align}
and consequently
\begin{align}
I_{\mathrm{true},n}(\theta;\mathbf y_{\mathrm{rc}}|\mathbf w_s)
=
I_{\mathrm{F\text{-}LMI},n}(\theta)
+
O \left(\kappa_n^{-1/2}\right).
\label{eq:flmi_true_mi_consistency}
\end{align}
\end{proposition}

\begin{IEEEproof}
Let $\Delta\theta_n\triangleq\theta-\mu_\theta\sim\mathcal N(0,v_n)$. Using the integral form of the first-order Taylor expansion of $\mathbf f$ around $\mu_\theta$,
\begin{align}
\mathbf f(\mu_\theta+\Delta\theta_n)
=
\mathbf f(\mu_\theta)
+
\mathbf J(\mu_\theta)\Delta\theta_n
+
\mathbf r(\Delta\theta_n),
\end{align}
where the Taylor remainder is
\begin{align}
\mathbf r(\delta)
=
\delta^2
\int_0^1
(1-t)
\mathbf f''(\mu_\theta+t\delta)
\,dt .
\end{align}
Since $\mu_\theta$ is known, subtracting the local mean $\mathbf f(\mu_\theta)$ is a bijection, and so are the parameter shift $\theta\mapsto\Delta\theta_n$, the real stacking $\mathcal R(\cdot)$, and the scaling by $\sqrt{\kappa_n}$; all of these preserve MI. After these operations, the nonlinear and linearized channels become
\begin{align}
\tilde{\mathbf y}_{\mathrm{true},n}^{\mathrm w}
=
\mathbf U_n+\mathbf V_n+\mathbf Z,
\qquad
\tilde{\mathbf y}_{\mathrm{lin},n}^{\mathrm w}
=
\mathbf U_n+\mathbf Z,
\end{align}
where $\mathbf Z\sim\mathcal N(\mathbf 0,\mathbf I_{2N_R})$ and
\begin{align}
\mathbf U_n
\triangleq
\sqrt{\kappa_n}\,
\mathcal R \left(\mathbf J(\mu_\theta)\Delta\theta_n\right),
\qquad
\mathbf V_n
\triangleq
\sqrt{\kappa_n}\,
\mathcal R \left(\mathbf r(\Delta\theta_n)\right).
\end{align}
Here the real-stacked noise covariance is $\kappa_n^{-1}\mathbf I_{2N_R}$, so multiplication by $\sqrt{\kappa_n}$ whitens it. Write $I_{\mathrm{true},n}\triangleq I(\Delta\theta_n;\tilde{\mathbf y}_{\mathrm{true},n}^{\mathrm w})$ and $I_{\mathrm{lin},n}\triangleq I(\Delta\theta_n;\tilde{\mathbf y}_{\mathrm{lin},n}^{\mathrm w})$.

Since the two channels differ only through $\mathbf V_n$, we first bound its size at the noise-whitened scale. By (A1), $\|\mathbf r(\delta)\|_2\le\frac{C_f}{2}|\delta|^2$, and since $\mathbb E|\Delta\theta_n|^4=3v_n^2$,
\begin{align}
\mathbb E\|\mathbf V_n\|_2^2
=
\kappa_n\,
\mathbb E\|\mathbf r(\Delta\theta_n)\|_2^2
\le
\frac{3C_f^2}{4}
\kappa_n v_n^2 .
\label{eq:remainder_second_moment}
\end{align}
By (A2)--(A3), $\kappa_n v_n=\rho_n/\|\mathbf J(\mu_\theta)\|_2^2=\Theta(1)$, hence $v_n=\Theta(\kappa_n^{-1})$ and
\begin{align}
\left(\mathbb E\|\mathbf V_n\|_2^2\right)^{1/2}
\le
\frac{\sqrt3\,C_f}{2}\sqrt{\kappa_n}\,v_n
=
\frac{\sqrt3\,C_f\,\rho_n}
{2\|\mathbf J(\mu_\theta)\|_2^2}
\,\kappa_n^{-1/2}.
\label{eq:derived_remainder_bound}
\end{align}
Rather than evaluating each mutual information separately, we interpolate between the two channels: for $t\in[0,1]$ define the interpolating channel $\mathbf Y_{t,n}\triangleq\mathbf U_n+t\mathbf V_n+\mathbf Z$, so that $\mathbf Y_{0,n}=\tilde{\mathbf y}_{\mathrm{lin},n}^{\mathrm w}$ and $\mathbf Y_{1,n}=\tilde{\mathbf y}_{\mathrm{true},n}^{\mathrm w}$. Because $\mathbf U_n$ and $\mathbf V_n$ are both deterministic functions of $\Delta\theta_n$, conditioning on $\Delta\theta_n$ leaves only the additive Gaussian noise, so $h(\mathbf Y_{t,n}\mid\Delta\theta_n)=h(\mathbf Z)$ for all $t\in[0,1]$; hence only the marginal entropy varies with $t$ and $I(\Delta\theta_n;\mathbf Y_{t,n})=h(\mathbf Y_{t,n})-h(\mathbf Z)$. Let $p_{t,n}$ denote the density of $\mathbf Y_{t,n}$, $\mathbf s_{t,n}(\mathbf y)\triangleq\nabla_{\mathbf y}\log p_{t,n}(\mathbf y)$, and $\mathbf m_{t,n}(\mathbf y)\triangleq\mathbb E[\mathbf V_n\mid\mathbf Y_{t,n}=\mathbf y]$. Since $p_{t,n}$ is a Gaussian mixture, it is smooth and strictly positive, and the uniform second-moment bounds established above justify differentiation under the integral sign and integration by parts. Writing $\phi$ for the standard Gaussian density and using $\nabla\phi(\mathbf u)=-\mathbf u\phi(\mathbf u)$, we obtain $\partial_t p_{t,n}=-\nabla \cdot \left(p_{t,n}\mathbf m_{t,n}\right)$, and therefore, in nats,
\begin{align}
\frac{d}{dt}h(\mathbf Y_{t,n})
=
-
\mathbb E
\left[
\mathbf m_{t,n}^{\sf T}(\mathbf Y_{t,n})
\mathbf s_{t,n}(\mathbf Y_{t,n})
\right].
\end{align}
The same Gaussian identity gives the Tweedie-type relation $\mathbf s_{t,n}(\mathbf Y_{t,n})=-\mathbb E[\mathbf Z\mid\mathbf Y_{t,n}]$, so that by conditional Jensen $\mathbb E\|\mathbf s_{t,n}(\mathbf Y_{t,n})\|_2^2\le\mathbb E\|\mathbf Z\|_2^2=2N_R$ and, likewise, $\mathbb E\|\mathbf m_{t,n}(\mathbf Y_{t,n})\|_2^2\le\mathbb E\|\mathbf V_n\|_2^2$. Applying Cauchy--Schwarz,
\begin{align}
\left|
\frac{d}{dt}h(\mathbf Y_{t,n})
\right|
\le
\left(
\mathbb E\|\mathbf V_n\|_2^2
\right)^{1/2}
\sqrt{2N_R},
\qquad t\in[0,1].
\end{align}
Integrating over $t\in[0,1]$ and converting from nats to bits,
\begin{align}
\left|
I_{\mathrm{true},n}
-
I_{\mathrm{lin},n}
\right|
\le
\frac{\sqrt{2N_R}}{\ln 2}
\left(
\mathbb E\|\mathbf V_n\|_2^2
\right)^{1/2}.
\label{eq:interpolation_bound}
\end{align}

It remains to evaluate the linearized MI. The channel $\mathbf Y_{0,n}=\sqrt{\kappa_n}\,\mathcal R(\mathbf J(\mu_\theta))\Delta\theta_n+\mathbf Z$ is a scalar-input real Gaussian channel with a rank-one signal covariance, so by the matrix determinant lemma and norm preservation under real stacking,
\begin{align}
I_{\mathrm{lin},n}
&=
\frac{1}{2}
\log_2
\det
\left(
\mathbf I_{2N_R}
+
\kappa_n v_n
\mathcal R(\mathbf J(\mu_\theta))
\mathcal R(\mathbf J(\mu_\theta))^{\sf T}
\right)
\nonumber\\
&=
\frac{1}{2}
\log_2
\left(
1+
\kappa_n v_n
\|\mathbf J(\mu_\theta)\|_2^2
\right)
=
I_{\mathrm{F\text{-}LMI},n}.
\end{align}
Substituting \eqref{eq:derived_remainder_bound} into \eqref{eq:interpolation_bound} and identifying $I_{\mathrm{lin},n}=I_{\mathrm{F\text{-}LMI},n}$ yields \eqref{eq:flmi_nonasymptotic_bound}, and \eqref{eq:flmi_true_mi_consistency} follows since $\rho_n=\Theta(1)$.
\end{IEEEproof}

% The value of Proposition~\ref{prop:flmi_consistency} is not that a concentrated prior makes the linearization accurate, which is expected, but that \eqref{eq:flmi_nonasymptotic_bound} identifies the scaling under which F-LMI remains informative. Two extremes clarify this point. If $v_n$ were held fixed while $\kappa_n\to\infty$, the angular deviations would stay large enough for the second-order term of the array response to matter, so the linearized model would no longer describe the true one. If instead $\rho_n\to0$, the measurement would carry almost no information about the angle, and both sides of \eqref{eq:flmi_true_mi_consistency} would vanish, leaving the statement empty.

Proposition~\ref{prop:flmi_consistency} identifies the scaling under
which F-LMI remains a meaningful information measure. If $v_n$ were fixed
while $\kappa_n\to\infty$, the second-order term would persist after noise
whitening and the linearized model would no longer describe the true one;
if $\rho_n\to0$, both sides of \eqref{eq:flmi_true_mi_consistency} would
vanish. The balanced regime $\rho_n=\Theta(1)$, motivated by tracking
updates with a finite information gain, avoids both extremes. The
proposition is thus a consistency guarantee for this regime rather than an
accuracy claim at a particular $L$ and $\sigma_\theta$; it justifies
replacing the intractable MI in \eqref{eq:mi_def} by a closed-form
bit-domain surrogate, whose designs are validated in
Section~\ref{sec:simulation}.

While Proposition~\ref{prop:flmi_consistency} relates F-LMI to the MI of the
nonlinear model, the following corollary re-expresses the same quantity as a
log-ratio between the prior variance and the Bayesian bound, thereby
connecting it to the FI/CRB family of Section~\ref{sec:preliminaries}.

\begin{corollary}[Local BCRB log-ratio interpretation]
\label{cor:flmi_bcrb}
Under the conditions of Proposition~\ref{prop:flmi_consistency}, let
\begin{align}
\mathrm{BCRB}^{\mathrm{true}}_n
\triangleq
\left(
v_n^{-1}
+
\mathbb E_{\theta}
\left[
\CMcal{J}^{\mathrm{true}}_n(\theta)
\right]
\right)^{-1}
\end{align}
denote the scalar BCRB of the nonlinear range-compressed model. Then, for every $n$,
\begin{align}
\left|
I_{\mathrm{F\text{-}LMI},n}(\theta)
-
\frac{1}{2}
\log_2
\left(
\frac{v_n}
{\mathrm{BCRB}^{\mathrm{true}}_n}
\right)
\right|
\le
\frac{c_n}{2\ln 2},
\label{eq:bcrb_nonasymptotic_bound}
\end{align}
where
\begin{align}
c_n
\triangleq
\kappa_n v_n
\left(
2\|\mathbf J(\mu_\theta)\|_2 C_f
\sqrt{\frac{2v_n}{\pi}}
+
C_f^2 v_n
\right)
=
O\left(\kappa_n^{-1/2}\right).
\end{align}
Consequently,
\begin{align}
I_{\mathrm{F\text{-}LMI},n}(\theta)
=
\frac{1}{2}
\log_2
\left(
\frac{v_n}
{\mathrm{BCRB}^{\mathrm{true}}_n}
\right)
+
O \left(\kappa_n^{-1/2}\right).
\label{eq:bcrb_logratio}
\end{align}
\end{corollary}

\begin{IEEEproof}
By (A1), for any $\delta$,
\begin{align}
\mathbf J(\mu_\theta+\delta)-\mathbf J(\mu_\theta)
=
\int_0^\delta
\mathbf f''(\mu_\theta+t)
\,dt ,
\end{align}
so that, writing $\mathbf e(\delta)\triangleq\mathbf J(\mu_\theta+\delta)-\mathbf J(\mu_\theta)$, we have $\|\mathbf e(\delta)\|_2\le C_f|\delta|$. Expanding the squared norm gives
\begin{align}
\|\mathbf J(\mu_\theta+\delta)\|_2^2
-
\|\mathbf J(\mu_\theta)\|_2^2
=
2\operatorname{Re}
\left\{
\mathbf J^{\sf H}(\mu_\theta)\mathbf e(\delta)
\right\}
+
\|\mathbf e(\delta)\|_2^2 ,
\end{align}
and therefore, by the Cauchy--Schwarz inequality,
\begin{align}
\left|
\|\mathbf J(\mu_\theta+\delta)\|_2^2
-
\|\mathbf J(\mu_\theta)\|_2^2
\right|
\le
2\|\mathbf J(\mu_\theta)\|_2 C_f|\delta|
+
C_f^2\delta^2 .
\end{align}
For $\Delta\theta_n\sim\mathcal N(0,v_n)$ we have $\mathbb E|\Delta\theta_n|=\sqrt{2v_n/\pi}$ and $\mathbb E|\Delta\theta_n|^2=v_n$. Since $\CMcal{J}^{\mathrm{true}}_n(\theta)=\kappa_n\|\mathbf J(\theta)\|_2^2$ by \eqref{eq:true_fi_def}, taking expectations gives
\begin{align}
v_n
\left|
\mathbb E_\theta
\left[
\CMcal{J}^{\mathrm{true}}_n(\theta)
\right]
-
\CMcal{J}^{\mathrm{true}}_n(\mu_\theta)
\right|
\le
c_n ,
\label{eq:fi_averaging_gap}
\end{align}
with $c_n$ as defined above. By (A2)--(A3), $\kappa_n v_n=\Theta(1)$ and $v_n=\Theta(\kappa_n^{-1})$, hence $c_n=O(\sqrt{v_n})=O(\kappa_n^{-1/2})$.

By the definition of $\mathrm{BCRB}^{\mathrm{true}}_n$,
\begin{align}
\frac{v_n}{\mathrm{BCRB}^{\mathrm{true}}_n}
=
1+
v_n
\mathbb E_\theta
\left[
\CMcal{J}^{\mathrm{true}}_n(\theta)
\right],
\end{align}
whereas $I_{\mathrm{F\text{-}LMI},n}(\theta)=\frac{1}{2}\log_2(1+v_n\CMcal{J}^{\mathrm{true}}_n(\mu_\theta))$. Both arguments of the logarithm are at least one, so the mean value theorem applied to $x\mapsto\log_2(1+x)$ bounds the difference of the two logarithms by $1/\ln 2$ times the difference of the arguments, which is exactly \eqref{eq:fi_averaging_gap}. This yields \eqref{eq:bcrb_nonasymptotic_bound}, and \eqref{eq:bcrb_logratio} follows since $c_n=O(\kappa_n^{-1/2})$.
\end{IEEEproof}

Within the stated local regime, F-LMI approximates the logarithmic
reduction from prior variance to the nonlinear Bayesian error bound
for the range-compressed observation. This is a pointwise statement
for a fixed beamformer, not a global optimizer-equivalence statement.
We now turn to the multi-target case, which is the main formulation
of this paper.

\subsection{Multi-Target Scenario}
\label{sec:flmi_multi}

For $M>1$, let $\boldsymbol{\theta}\triangleq[\theta_1,\dots,\theta_M]^{\sf T}$ denote the unknown target directions, and assume a Gaussian prior $\boldsymbol{\theta}\sim\mathcal{N}(\boldsymbol{\mu}_\theta,\mathbf{\Sigma}_\theta)$, where $\mathbf{\Sigma}_\theta = \mathrm{diag}(\sigma_{\theta_1}^2, \dots, \sigma_{\theta_M}^2)$.
Recalling the sensing model \eqref{eq:radar_obs_model}, let $\mathbf{s}_{s,m}\in\mathbb{C}^{1\times L}$ denote the $m$-th row of $\mathbf{S}_s$, and $\mathbf{w}_{s,m}\in\mathbb{C}^{N_T\times 1}$ denote the $m$-th column of $\mathbf{W}_s$.
We associate each target-centric sensing stream with a dedicated sequence $\mathbf s_{s,m}$, which enables separation by range compression, whereas $\mathbf w_{s,m}$ provides spatial selectivity. Let $\mathbf S\triangleq[\mathbf S_c^{\sf T},\mathbf S_s^{\sf T}]^{\sf T} \in\mathbb C^{(K+M)\times L}$ have full row rank ($L\ge K+M$), so that $\mathbf X=[\mathbf W_c,\mathbf W_s]\mathbf S$. As in \eqref{eq:single_target_model_mf}, the range-compressed statistic for the $m$-th sensing stream is
\begin{align}
\mathbf y_{\mathrm{rc},m}
&\triangleq
\mathbf Y_R\mathbf S^{\sf H}(\mathbf S\mathbf S^{\sf H})^{-1}\mathbf e_{K+m}
\nonumber\\
&=\sum_{\ell=1}^{M}\mathbf A(\theta_\ell)\mathbf w_{s,m}+\bar{\mathbf z}_m,
\label{eq:rc_model_multi}
\end{align}
where, conditional on $\mathbf S$, $\bar{\mathbf z}_m\sim\mathcal{CN}(\mathbf 0,\sigma_{\mathrm{rc},m}^2\mathbf I_{N_R})$ with $\sigma_{\mathrm{rc},m}^2\triangleq \sigma_z^2[(\mathbf S\mathbf S^{\sf H})^{-1}]_{K+m,K+m}$. The retained outputs contain neither communication-stream nor cross-stream sensing leakage. The output noises may be correlated across branches; each branch-wise utility uses the corresponding marginal variance.
% Since the observation in \eqref{eq:rc_model_multi} involves a superposition of non-linear array responses from all targets, exact MI derivation is intractable.
Following the single-target case, we apply the first-order expansion $\mathbf A(\theta_\ell)\approx\mathbf{A}(\mu_{\theta_\ell}) +\dot{\mathbf{A}}(\mu_{\theta_\ell})(\theta_\ell-\mu_{\theta_\ell})$ around each prior mean. Defining $\Delta\theta_\ell\triangleq\theta_\ell-\mu_{\theta_\ell}$, we obtain the centered range-compressed observation
\begin{align}
\bar{\mathbf y}_{\mathrm{rc},m}
\triangleq
\mathbf y_{\mathrm{rc},m}
-\sum_{\ell=1}^{M}
\mathbf A(\mu_{\theta_\ell})\mathbf w_{s,m}
\approx \sum_{\ell=1}^{M}\mathbf{J}_{\ell,m}\Delta\theta_\ell+\bar{\mathbf{z}}_{m},
\label{eq:rc_model_multi_marginal}
\end{align}
where $\mathbf{J}_{\ell,m}\triangleq \dot{\mathbf{A}}(\mu_{\theta_\ell})\mathbf{w}_{s,m}\in\mathbb{C}^{N_R\times 1}$.
To rigorously handle the improper statistics caused by real-valued parameters $\Delta\theta_\ell$, we apply the real-valued isomorphism in \eqref{eq:real_vectorization} to the $m$-th statistic.
Applying the real-valued operators \(\mathcal R(\cdot)\) and
\(\mathcal T(\cdot)\) to \eqref{eq:rc_model_multi_marginal},
define
\begin{align}
\tilde{\mathbf J}_{\ell,m}
\triangleq
\mathcal T \left(\dot{\mathbf A}(\mu_{\theta_\ell})\right)
\mathcal R(\mathbf w_{s,m})
\in\mathbb R^{2N_R}.
\end{align}
The associated real Gaussian surrogate is defined by
\begin{align}
\tilde{\mathbf y}_{\mathrm{eff},m}
\triangleq
\tilde{\mathbf J}_{m,m}\Delta\theta_m
+
\sum_{\ell\ne m}
\tilde{\mathbf J}_{\ell,m}\Delta\theta_\ell
+
\tilde{\mathbf z}_m,
\label{eq:real_model_explicit}
\end{align}
where $\tilde{\mathbf{z}}_{m} \sim \mathcal{N}(\mathbf{0}, \frac{\sigma_{\mathrm{rc},m}^2}{2}\mathbf{I}_{2N_R})$.

By treating the inter-target interference term in 
\eqref{eq:real_model_explicit} as structured interference, we derive the marginal MI $I(\tilde{\mathbf{y}}_{\text{eff},m};\theta_m|\mathbf{W}_{s})$ for the $m$-th target. 
The effective interference-plus-noise covariance matrix $\mathbf R_{\mathrm{in},m}$ is derived as
\begin{align}
\mathbf R_{\mathrm{in},m} = \sum_{\ell \neq m} \sigma_{\theta_\ell}^2 \tilde{\mathbf{J}}_{\ell,m}\tilde{\mathbf{J}}_{\ell,m}^{\sf T} + \frac{\sigma_{\mathrm{rc},m}^2}{2}\mathbf{I}_{2N_R}.
\label{eq:rin_def}
\end{align}
Using this covariance, the marginal MI admits the compact closed form
\begin{align}
I(\tilde{\mathbf{y}}_{\text{eff},m};\theta_m|\mathbf{W}_{s})
&= \frac{1}{2}\log_2 \left(
1 + \sigma_{\theta_m}^2 \tilde{\mathbf{J}}_{m,m}^{\sf T} \mathbf R_{\mathrm{in},m}^{-1} \tilde{\mathbf{J}}_{m,m}
\right),
\label{eq:flmi_marginal_compact}
\end{align}
which we denote by $I(\theta_m)$.

The quadratic form
$q_m\triangleq\tilde{\mathbf J}_{m,m}^{\sf T}\mathbf R_{\mathrm{in},m}^{-1}\tilde{\mathbf J}_{m,m}$
in \eqref{eq:flmi_marginal_compact} determines the local posterior
variance $v_m^{\mathrm{post}}=(\sigma_{\theta_m}^{-2}+q_m)^{-1}$ of
$\Delta\theta_m$ given the $m$-th output, so that
$I(\theta_m)=\frac12\log_2(\sigma_{\theta_m}^2/v_m^{\mathrm{post}})$ is the
Bayesian D-optimal information gain of this branch. The quadratic form $q_m$ equals
$\|\mathbf R_{\mathrm{in},m}^{-1/2}\tilde{\mathbf J}_{m,m}\|_2^2$ and is
upper-bounded by the noise-only term
$(2/\sigma_{\mathrm{rc},m}^2)\|\tilde{\mathbf J}_{m,m}\|_2^2$, with
equality if and only if $\tilde{\mathbf J}_{m,m}$ is orthogonal to every
nuisance sensitivity $\tilde{\mathbf J}_{\ell,m}$, $\ell\neq m$. A beam is
therefore rewarded not simply for illuminating the prior-mean direction,
but for producing intended angular changes that remain distinguishable
from the prior-induced variations of the other targets. Summing these
local marginal information terms across the assigned sensing outputs
yields
\begin{align}
I_{\mathrm{F\text{-}LMI}}(\boldsymbol{\theta})
\triangleq
\sum_{m=1}^{M} I(\theta_m).
\label{eq:multi_target_flmi}
\end{align}
Under independent angular priors, the sum in \eqref{eq:multi_target_flmi} is upper-bounded by the joint MI of all retained outputs under the same local Gaussian model. It measures information available to assigned-output target updates. Exhaustive nonlinear joint search or tensor-product integration considers $G^M$ angle tuples for $G$ candidates per angle. Gaussian nuisance modeling instead permits one-dimensional branch inference, totaling $MG$ intended-angle evaluations after forming the nuisance statistics. Neighboring echoes remain in each output; the reduction combines branch processing with statistical nuisance modeling. Each term is in bits per block and is normalized by $L$ in Section~\ref{sec:optimization}.

The separable form of~\eqref{eq:multi_target_flmi} also enables explicit target-priority control. Assigning a nonnegative weight $\rho_m$ to each target, we define the weighted sensing utility as
\begin{align}
R_S^{\mathrm{F\text{-}LMI}}(\mathbf{W}_s) 
\triangleq 
\sum_{m=1}^{M} \rho_m I(\theta_m) 
= 
\sum_{m=1}^{M} \frac{\rho_m}{2} \log_2 \left(1+\sigma_{\theta_m}^2 q_m\right),
\label{eq:weighted_sum_objective}
\end{align}
where $q_m \triangleq \tilde{\mathbf{J}}_{m,m}^{\sf T}\mathbf{R}_{\mathrm{in},m}^{-1}\tilde{\mathbf{J}}_{m,m}$, so that the beamformer can reallocate sensing resources across targets according to their priorities.

F-LMI guides beam design through local angular sensitivities
and nuisance covariance. The operational receivers retain the
nonlinear intended response and can exploit its curvature during
inference. The rate--RMSE experiments therefore evaluate the
resulting beams under nonlinear echoes and finite prior widths,
complementing the fixed-beamformer local consistency analysis.

\section{F-LMI-Based Beamforming Design}
\label{sec:optimization}

We develop an FP-based alternating algorithm for the joint
multi-target communication--sensing beamforming problem
\cite{ShenTWC18,ShenToN19_MatrixFP,ChenTWC25_FastFP_FI}.

\subsection{Joint Utility and Real-Valued F-LMI Reformulation}
\label{subsec:joint_flmi_reformulation}

The key difficulty is that the proposed sensing utility is a logarithmic local-MI metric rather than a conventional quadratic FI objective. Moreover, the interference-plus-noise covariance associated with each target-specific range-compressed statistic depends on the sensing beamformer being optimized, since the sensitivities to the non-intended targets are also shaped by the same sensing stream. This yields a coupled log-fractional structure that cannot be handled by a scalar FP transform alone. To address this, we show that, after an equivalent real-valued reformulation, the receiver-oriented F-LMI can be optimized by combining the Lagrangian dual transform with scalar and matrix quadratic transforms. With fixed auxiliary variables, the beamformer update reduces to closed-form linear-system updates, together with a scalar bisection for the total transmit-power constraint.

\subsubsection{Weighted Utility and Normalization}

Recall that the communication rate is
\(R_k=\log_2(1+\gamma_{c,k})\), and, from
\eqref{eq:flmi_marginal_compact} and
\eqref{eq:weighted_sum_objective}, the sensing utility of the
\(m\)-th range-compressed statistic can be written as
\(I_m\triangleq I(\theta_m)=\frac{1}{2}\log_2(1+\gamma_{s,m})\), where
\(\gamma_{s,m}\triangleq\sigma_{\theta_m}^2 q_m\) is made explicit in
\eqref{eq:gamma_s_real_fp} below. For data-independent design, the sensing covariances and SINRs below are evaluated using the mean compressed-noise variance
$\bar{\sigma}_{\mathrm{rc}}^2
\triangleq\mathbb E_{\mathbf S}[\sigma_{\mathrm{rc},m}^2]
=\sigma_z^2/(L-K-M)$, valid for i.i.d.\ $\mathcal{CN}(0,1)$
symbols and $L>K+M$ \cite{Maiwald2000_InvWishart}. For fixed precoders, this gives a Jensen lower bound on the waveform-averaged local Gaussian branch utility; the receiver uses the actual conditional variances. We use effective nonnegative weights
\(a_{c,k}\) and \(b_{s,m}\) and solve
\begin{subequations}
\label{eq:joint_opt_effective}
\begin{align}
\max_{\mathbf W_c,\mathbf W_s}\quad
&
\sum_{k=1}^{K}
a_{c,k}\log_2(1+\gamma_{c,k})
+
\frac{1}{L}\sum_{m=1}^{M}
\frac{b_{s,m}}{2}\log_2(1+\gamma_{s,m})
\label{eq:joint_obj}\\
&
\mathrm{s.t.}\quad
\|\mathbf W_c\|_F^2+\|\mathbf W_s\|_F^2\le P_{\mathrm{tx}}.
\label{eq:joint_const}
\end{align}
\end{subequations}
The factor $1/L$ expresses the sensing utility per channel use. For compact notation, let $\lambda_{c,k}=a_{c,k}/\ln2$ and $\lambda_{s,m}=b_{s,m}/(2L\ln2)$.

\subsubsection{Real-Valued Representation of Target-Specific Statistics}

Using the real-valued operators \(\mathcal R(\cdot)\) and
\(\mathcal T(\cdot)\) defined in Section~\ref{sec:flmi}, let
\(\tilde{\mathbf w}_{s,m}=\mathcal R(\mathbf w_{s,m})\),
\(\mathbf B_\ell=\mathcal T(\dot{\mathbf A}(\mu_{\theta_\ell}))\), and
\(\mathbf j_{\ell,m}=\mathbf B_\ell\tilde{\mathbf w}_{s,m}\), i.e.,
the real-stacked sensitivity \(\tilde{\mathbf J}_{\ell,m}\) of the
\(m\)-th statistic to the \(\ell\)-th target angle. The desired local-sensitivity vector for target \(m\) is
\begin{align}
\mathbf s_m
\triangleq
\sigma_{\theta_m}\mathbf j_{m,m}
=
\sigma_{\theta_m}\mathbf B_m\tilde{\mathbf w}_{s,m}.
\end{align}
The corresponding interference-plus-noise covariance $\mathbf R_{\mathrm{in},m}$ is given by \eqref{eq:rin_def} with $\tilde{\mathbf{J}}_{\ell,m}=\mathbf j_{\ell,m}$ and $\sigma_{\mathrm{rc},m}^2$ replaced by $\bar\sigma_{\mathrm{rc}}^2$.
Accordingly, the F-LMI sensing SINR can be written as
\begin{align}
\gamma_{s,m}
=
\mathbf s_m^{\sf T}
\mathbf R_{\mathrm{in},m}^{-1}
\mathbf s_m.
\label{eq:gamma_s_real_fp}
\end{align}

\subsubsection{Matrix-FP-Compatible F-LMI Fraction}

Define the total covariance
\begin{align}
\mathbf R_{y,m}
=
\mathbf R_{\mathrm{in},m}
+
\mathbf s_m\mathbf s_m^{\sf T}.
\end{align}
Since \(\bar\sigma_{\mathrm{rc}}^2>0\), both
\(\mathbf R_{\mathrm{in},m}\) and \(\mathbf R_{y,m}\) are positive
definite. By the matrix inversion lemma,
\begin{align}
\frac{\gamma_{s,m}}{1+\gamma_{s,m}}
=
\mathbf s_m^{\sf T}\mathbf R_{y,m}^{-1}\mathbf s_m .
\label{eq:woodbury_gamma_ratio}
\end{align}
The relevance of \eqref{eq:woodbury_gamma_ratio} is that
$\gamma_{s,m}/(1+\gamma_{s,m})$ is exactly the sensing ratio that appears
after the Lagrangian-dual transform in
Section~\ref{subsec:fp_optimization}; the identity converts it into a
covariance-weighted quadratic form compatible with the matrix quadratic
transform. Specifically, for any $\chi_m>0$,
\begin{align}
\chi_m
\mathbf s_m^{\sf T}\mathbf R_{y,m}^{-1}\mathbf s_m
=
\max_{\mathbf v_m}
\left[
2\sqrt{\chi_m}\,\mathbf v_m^{\sf T}\mathbf s_m
-
\mathbf v_m^{\sf T}\mathbf R_{y,m}\mathbf v_m
\right].
\label{eq:matrix_qt_preview}
\end{align}
In the subsequent Lagrangian-dual step, \(\chi_m=1+u_{s,m}\). For fixed
\(\mathbf v_m\), the transformed sensing term is a concave quadratic
function of the real-stacked sensing beamformer because
\(\mathbf s_m\) is linear in \(\tilde{\mathbf w}_{s,m}\) and
\(\mathbf R_{y,m}\) is quadratic in
\(\{\mathbf B_\ell\tilde{\mathbf w}_{s,m}\}_{\ell=1}^{M}\).

\subsection{FP-Based Alternating Beamformer Optimization}
\label{subsec:fp_optimization}

\subsubsection{Lagrangian Dual Transform}

We first apply the Lagrangian dual transform to the logarithmic terms in
\eqref{eq:joint_opt_effective}. Introduce nonnegative scalar auxiliary
variables $\{u_{c,k}\}_{k=1}^{K}$ and $\{u_{s,m}\}_{m=1}^{M}$.
For fixed beamformers, their optimal updates are
\begin{align}
u_{c,k}^{\star}
=
\gamma_{c,k},
\qquad
u_{s,m}^{\star}
=
\gamma_{s,m}.
\label{eq:u_update_multi}
\end{align}
Using the Lagrangian dual transform, the objective in
\eqref{eq:joint_opt_effective} is equivalently represented as
\begin{align}
\mathcal F_{\mathrm{LDT}}
&=
\sum_{k=1}^{K}
\lambda_{c,k}
\left[
\ln(1+u_{c,k})-u_{c,k}
+
(1+u_{c,k})
\frac{\gamma_{c,k}}{1+\gamma_{c,k}}
\right]
\nonumber\\
&\quad+
\sum_{m=1}^{M}
\lambda_{s,m}
\left[
\ln(1+u_{s,m})-u_{s,m}
+
(1+u_{s,m})
\frac{\gamma_{s,m}}{1+\gamma_{s,m}}
\right].
\label{eq:ldt_full_surrogate}
\end{align}
For fixed auxiliary variables, the terms
$\ln(1+u_{c,k})-u_{c,k}$ and
$\ln(1+u_{s,m})-u_{s,m}$ are independent of the beamformers.
Therefore, the beamformer-dependent part is
\begin{align}
\bar{\mathcal F}_{\mathrm{LDT}}
&=
\sum_{k=1}^{K}
\lambda_{c,k}(1+u_{c,k})
\frac{\gamma_{c,k}}{1+\gamma_{c,k}}
\nonumber\\
&\quad+
\sum_{m=1}^{M}
\lambda_{s,m}(1+u_{s,m})
\frac{\gamma_{s,m}}{1+\gamma_{s,m}}
+
\mathrm{const}.
\label{eq:ldt_beamformer_dependent}
\end{align}
The factors $\sqrt{1+u_{c,k}}$ and $\sqrt{1+u_{s,m}}$ appearing in the
subsequent quadratic-transform updates originate from the coefficients
$(1+u_{c,k})$ and $(1+u_{s,m})$ in
\eqref{eq:ldt_beamformer_dependent}.

After the Lagrangian-dual transform, the communication terms contain the
standard ratios \(\gamma_{c,k}/(1+\gamma_{c,k})\), while the sensing terms
contain the covariance form in \eqref{eq:woodbury_gamma_ratio}. These two
types of ratios are handled by the scalar and matrix quadratic transforms,
respectively.

\subsubsection{Scalar and Matrix Quadratic Transforms}

For the communication terms, define
\begin{align}
D_{c,k}
\triangleq
\sum_{j=1}^{K}|\mathbf h_k^{\sf H}\mathbf w_{c,j}|^2
+
\sum_{m=1}^{M}|\mathbf h_k^{\sf H}\mathbf w_{s,m}|^2
+
\sigma_c^2.
\end{align}
Applying the standard scalar quadratic transform to the communication
ratio \(\gamma_{c,k}/(1+\gamma_{c,k})\) introduces auxiliary variables
\(\{\beta_{c,k}\}_{k=1}^{K}\). For fixed beamformers, the optimal update is
\begin{align}
\beta_{c,k}^{\star}
=
\frac{
\sqrt{1+u_{c,k}}\,
\mathbf h_k^{\sf H}\mathbf w_{c,k}
}
{D_{c,k}}.
\label{eq:beta_update_multi_final}
\end{align}
With $\{u_{c,k}\}$ and $\{\beta_{c,k}\}$ fixed, the communication surrogate contributes
\begin{align}
\sum_{k=1}^{K}
\left(
2\Re\{\mathbf d_{c,k}^{\sf H}\mathbf w_{c,k}\}
-
\mathbf w_{c,k}^{\sf H}\mathbf \Phi_c\mathbf w_{c,k}
\right)
-
\sum_{m=1}^{M}
\mathbf w_{s,m}^{\sf H}\mathbf \Phi_c\mathbf w_{s,m}
+
\mathrm{const},
\end{align}
where
\begin{align}
\mathbf \Phi_c
&\triangleq
\sum_{i=1}^{K}
\lambda_{c,i}|\beta_{c,i}|^2\mathbf h_i\mathbf h_i^{\sf H},
\label{eq:Phi_c_final}
\\
\mathbf d_{c,k}
&\triangleq
\lambda_{c,k}\sqrt{1+u_{c,k}}\,
\beta_{c,k}\mathbf h_k.
\label{eq:d_c_final}
\end{align}
The last negative term appears because the sensing streams are treated as interference at the communication users.

For the sensing term of the $m$-th target-specific statistic, we set
$\chi_m=1+u_{s,m}$ in \eqref{eq:matrix_qt_preview}. The first-order
optimality condition of \eqref{eq:matrix_qt_preview} with respect to
$\mathbf v_m$ then yields the optimal matrix-FP auxiliary vector
\begin{align}
\mathbf v_m^{\star}
=
\mathbf R_{y,m}^{-1}
\left(
\sqrt{1+u_{s,m}}\,
\sigma_{\theta_m}\mathbf B_m\tilde{\mathbf w}_{s,m}
\right).
\label{eq:v_update_multi_final}
\end{align}
After each auxiliary update, we overwrite
$\beta_{c,k}\leftarrow\beta_{c,k}^{\star}$ and
$\mathbf v_m\leftarrow\mathbf v_m^{\star}$, and omit the superscript
$\star$ in the subsequent surrogate coefficients for notational
simplicity.
For fixed $\mathbf v_m$, the covariance term in
\eqref{eq:matrix_qt_preview} expands as
\begin{align}
\mathbf v_m^{\sf T}\mathbf R_{y,m}\mathbf v_m
&=
\sum_{\ell=1}^{M}
\sigma_{\theta_\ell}^{2}
\left(
\mathbf v_m^{\sf T}\mathbf B_\ell\tilde{\mathbf w}_{s,m}
\right)^2
+
\frac{\bar\sigma_{\mathrm{rc}}^2}{2}\|\mathbf v_m\|_2^2
\nonumber\\
&=
\tilde{\mathbf w}_{s,m}^{\sf T}
\left(
\sum_{\ell=1}^{M}
\sigma_{\theta_\ell}^{2}
\mathbf B_\ell^{\sf T}
\mathbf v_m\mathbf v_m^{\sf T}
\mathbf B_\ell
\right)
\tilde{\mathbf w}_{s,m}
+
\mathrm{const}.
\label{eq:Ry_quadratic_expansion}
\end{align}
The noise term in \eqref{eq:Ry_quadratic_expansion} is independent of
$\tilde{\mathbf w}_{s,m}$ and is therefore absorbed into the constant term.
Restoring the sensing weight $\lambda_{s,m}$ from
\eqref{eq:ldt_beamformer_dependent} and combining the linear term in
\eqref{eq:matrix_qt_preview} with
\eqref{eq:Ry_quadratic_expansion}, the sensing-side contribution to the
surrogate becomes
\begin{align}
2\mathbf d_{s,m}^{\sf T}\tilde{\mathbf w}_{s,m}
-
\tilde{\mathbf w}_{s,m}^{\sf T}
\mathbf \Psi_m
\tilde{\mathbf w}_{s,m}
+
\mathrm{const},
\end{align}
where
\begin{align}
\mathbf d_{s,m}
&\triangleq
\lambda_{s,m}
\sqrt{1+u_{s,m}}\,
\sigma_{\theta_m}\mathbf B_m^{\sf T}\mathbf v_m,
\label{eq:d_s_final}
\\
\mathbf \Psi_m
&\triangleq
\lambda_{s,m}
\sum_{\ell=1}^{M}
\sigma_{\theta_\ell}^2
\mathbf B_\ell^{\sf T}
\mathbf v_m\mathbf v_m^{\sf T}
\mathbf B_\ell .
\label{eq:Psi_final}
\end{align}
The summation in \eqref{eq:Psi_final} includes all targets, including the intended one, because it comes from the total covariance $\mathbf R_{y,m}$ rather than only from the interference covariance $\mathbf R_{\mathrm{in},m}$.

\subsubsection{Total-Power Beamformer Update}

To combine the complex communication penalty with the real-stacked sensing update, let \(\tilde{\mathbf \Phi}_c\triangleq\mathcal T(\mathbf \Phi_c)\).
Collecting the communication and sensing terms, the fixed-auxiliary beamformer subproblem is
\begin{subequations}
\label{eq:fixed_aux_quad_subproblem}
\begin{align}
\max_{\mathbf W_c,\{\tilde{\mathbf w}_{s,m}\}}
\quad
&
\sum_{k=1}^{K}
\left(
2\Re\{\mathbf d_{c,k}^{\sf H}\mathbf w_{c,k}\}
-
\mathbf w_{c,k}^{\sf H}\mathbf \Phi_c\mathbf w_{c,k}
\right)
\nonumber\\
&+
\sum_{m=1}^{M}
\left(
2\mathbf d_{s,m}^{\sf T}\tilde{\mathbf w}_{s,m}
-
\tilde{\mathbf w}_{s,m}^{\sf T}
(\tilde{\mathbf \Phi}_c+\mathbf \Psi_m)
\tilde{\mathbf w}_{s,m}
\right)
\\
\mathrm{s.t.}\quad
&
\sum_{k=1}^{K}\|\mathbf w_{c,k}\|_2^2
+
\sum_{m=1}^{M}\|\tilde{\mathbf w}_{s,m}\|_2^2
\le
P_{\mathrm{tx}}.
\end{align}
\end{subequations}
For fixed auxiliary variables, \eqref{eq:fixed_aux_quad_subproblem} is a concave quadratic maximization over the total-power ball. Its Karush–Kuhn–Tucker (KKT) conditions yield
\begin{align}
\mathbf w_{c,k}^{\star}(\mu)
=
(\mathbf \Phi_c+\mu\mathbf I_{N_T})^{-1}\mathbf d_{c,k},
\qquad
k=1,\ldots,K,
\label{eq:wc_update_multi_final}
\end{align}
and
\begin{align}
\tilde{\mathbf w}_{s,m}^{\star}(\mu)
=
(\tilde{\mathbf \Phi}_c+\mathbf \Psi_m+\mu\mathbf I_{2N_T})^{-1}
\mathbf d_{s,m},
\qquad
m=1,\ldots,M.
\label{eq:ws_update_multi_final}
\end{align}
The complex sensing beamformer is recovered by the inverse real-stacking operation,
\begin{align}
\mathbf w_{s,m}^{\star}
=
\mathcal R^{-1}(\tilde{\mathbf w}_{s,m}^{\star}).
\end{align}

The dual variable $\mu\ge0$ is chosen to satisfy the total-power constraint. Define
\begin{align}
p(\mu)
&\triangleq
\sum_{k=1}^{K}
\left\|
(\mathbf \Phi_c+\mu\mathbf I_{N_T})^{-1}\mathbf d_{c,k}
\right\|_2^2
\nonumber\\
&\quad+
\sum_{m=1}^{M}
\left\|
(\tilde{\mathbf \Phi}_c+\mathbf \Psi_m+\mu\mathbf I_{2N_T})^{-1}
\mathbf d_{s,m}
\right\|_2^2 .
\label{eq:power_function_mu}
\end{align}
Since $\mathbf \Phi_c\succeq\mathbf 0$,
$\tilde{\mathbf \Phi}_c\succeq\mathbf 0$, and
$\mathbf \Psi_m\succeq\mathbf 0$, \(p(\mu)\) is non-increasing in
\(\mu\). Hence, if \(p(0)>P_{\mathrm{tx}}\), the unique
\(\mu\) satisfying \(p(\mu)=P_{\mathrm{tx}}\) can be found by
one-dimensional bisection. If \(p(0)\le P_{\mathrm{tx}}\), the KKT
solution is obtained with \(\mu=0\).\footnote{If a coefficient matrix at
\(\mu=0\) is singular, the inverse in
\eqref{eq:wc_update_multi_final}--\eqref{eq:ws_update_multi_final} is
interpreted as the minimum-norm Moore--Penrose solution whenever the
corresponding linear term lies in its range; otherwise, the active-power
case with \(\mu>0\) applies.}

\subsubsection{Algorithm}

The resulting alternating FP algorithm is summarized in Algorithm~\ref{alg:fp_flmi}. At each iteration, the auxiliary variables are updated in closed form, and the beamformer block is obtained from \eqref{eq:wc_update_multi_final}--\eqref{eq:ws_update_multi_final} with scalar bisection on $\mu$ when required.

The algorithm only requires a feasible initialization. In the simulations,
we initialize the sensing beams using the prior-mean steering vectors and
initialize the communication beams by a feasible random or regularized zero-forcing-type
precoder, followed by normalization to satisfy the total-power constraint.

\begin{algorithm}[t]
\caption{Alternating FP for Joint F-LMI Beamforming}
\label{alg:fp_flmi}
\small
\KwIn{Channel $\mathbf H$; target priors
$\{\mu_{\theta_m},\sigma_{\theta_m}^2\}_{m=1}^{M}$;
mean compressed-noise variance $\bar\sigma_{\mathrm{rc}}^2$;
effective weights $\{a_{c,k},b_{s,m}\}$; power budget
$P_{\mathrm{tx}}$; tolerance $\epsilon$}
\KwOut{Communication and sensing beamformers $\mathbf W_c,\mathbf W_s$}

Initialize $\mathbf W_c,\mathbf W_s$ such that
$\|\mathbf W_c\|_F^2+\|\mathbf W_s\|_F^2\le P_{\mathrm{tx}}$\;

\Repeat{relative utility increase $<\epsilon$}{
    Compute $\{\gamma_{c,k}\}$ and $\{\gamma_{s,m}\}$\;

    Update $\{u_{c,k}\}$, $\{u_{s,m}\}$,
    $\{\beta_{c,k}\}$, $\mathbf\Phi_c$, and
    $\{\mathbf d_{c,k}\}$ by
    \eqref{eq:u_update_multi}--\eqref{eq:d_c_final}\;

    Form $\{\mathbf R_{y,m}\}$ and update
    $\{\mathbf v_m\}$, $\{\mathbf d_{s,m}\}$, and
    $\{\mathbf\Psi_m\}$ by
    \eqref{eq:v_update_multi_final}--\eqref{eq:Psi_final}\;

    \eIf{$p(0)\le P_{\mathrm{tx}}$}{
        Set $\mu^\star=0$\;
    }{
        Find $\mu^\star>0$ by bisection using
    \eqref{eq:power_function_mu}\;
    }

    Update $\mathbf W_c$ and $\mathbf W_s$ by
    \eqref{eq:wc_update_multi_final}--\eqref{eq:ws_update_multi_final}\;
}
\end{algorithm}

\subsection{Convergence and Complexity Analysis}
\label{subsec:conv_complexity}

The proposed algorithm alternates over the Lagrangian-dual variables,
quadratic-transform variables, and beamformers. For fixed beamformers,
the auxiliary updates are globally optimal for their FP surrogate
functions; for fixed auxiliary variables, the beamformer update solves the
concave quadratic subproblem under the total-power constraint. Hence, the mean-noise design objective in \eqref{eq:joint_opt_effective} is non-decreasing across iterations and, being bounded on the compact feasible set, converges.

Ignoring constant factors due to real stacking, the dominant
per-iteration complexity is $\mathcal O \left(
M N_R^3+(M+1)N_T^3+K(K+M)N_T
\right)$,
with an additional logarithmic factor in the bisection accuracy.

\section{Simulation Results}
\label{sec:simulation}

\subsection{Simulation Setup}

Unless otherwise stated, we use the default parameters in
Table~\ref{tab:sim_params}. The BS-to-user channels follow
the one-ring spatial-correlation model
\cite{ShiuTCOM2000_OneRing}, and perfect channel state
information at the transmitter (CSIT) is assumed. For sensing,
we use the monostatic point-target model in
Section~\ref{sec:system_model} with half-wavelength ULAs and
unit-norm steering vectors. The reference sensing SNR is
defined before range compression as
$\mathrm{SNR}_{r,0}=|\alpha_m|^2P_{\rm tx}/\sigma_z^2$
with $|\alpha_m|=1$, while $L$ enters through the
compressed-noise variance. The Gaussian angular priors
represent a tracking-oriented stage with coarse target
locations available from previous sensing intervals. We evaluate performance through Monte Carlo simulations over communication-channel realizations, with repeated sensing trials for each channel.

\begin{table}[t]
\caption{Default simulation parameters.}
\label{tab:sim_params}
\centering
\renewcommand{\arraystretch}{1.08}
\setlength{\tabcolsep}{4pt}
\begin{tabular}{l c}
\toprule
Parameter & Value \\
\midrule
\multicolumn{2}{l}{\textit{System and communication}} \\
BS transmit antennas, $N_T$ & $64$ \\
Sensing receive antennas, $N_R$ & $64$ \\
Number of users, $K$ & $16$ \\
Total transmit power, $P_{\mathrm{tx}}$ & $1$ \\
Communication noise variance, $\sigma_c^2$ & $0.1$ \\
User nominal directions, $\phi_k$ & $\mathcal U[-\pi/2,\pi/2]$ \\
One-ring angular spread, $\Delta$ & $10^\circ$ \\
\midrule
\multicolumn{2}{l}{\textit{Sensing}} \\
Number of targets, $M$ & $3$ \\
Pulse length, $L$ & $1000$ \\
Reflection coefficients, $\alpha_m$ & $1$ \\
Reference sensing SNR & $5$ dB \\
Angular prior std., $\sigma_{\theta_m}$ & $0.5^\circ$ \\
Well-separated layout & $\{-40^\circ,0^\circ,40^\circ\}$ \\
Closely spaced layout & $\{-3^\circ,0^\circ,3^\circ\}$ \\
\midrule
\multicolumn{2}{l}{\textit{Evaluation}} \\
Intended-angle window
(Figs.~\ref{fig:traj_close}--\ref{fig:sepsweep})
& $\pm3\sigma_{\theta_m}$ \\
Intended-angle window (Fig.~\ref{fig:common_basis})
& $\pm6\sigma_{\theta_m}$ \\
Nuisance integration window
(Figs.~\ref{fig:traj_close}--\ref{fig:traj_well})
& $\pm6\sigma_{\theta_\ell}$ \\
Angular grid step & $0.03^\circ$ unless otherwise stated \\
\bottomrule
\end{tabular}
\end{table}
\subsection{Baselines and Evaluation Protocol}
\label{subsec:evaluation_protocol}

% We compare the proposed F-LMI design with 
The main comparisons use four sensing baselines drawn from the two families reviewed in Section~\ref{sec:preliminaries}. 
Specifically, we consider Aggregate SMI and Joint BCRB adapted to the retained sensing outputs, together with two target-indexed (TI) variants, TI-SMI and TI-BCRB, in which the $m$-th sensing stream is tied to the $m$-th target so that the utility decomposes into per-target terms. 
% : the two conventional criteria, and the target-indexed variants that we construct to match the sensing-stream architecture of F-LMI.
All designs maximize a weighted sum of the communication sum-rate and the corresponding sensing utility under the same total-power constraint \eqref{eq:joint_const}, using $M$ dedicated sensing streams, one per target. 
The TI baselines use projected-gradient ascent (PGA) with Armijo backtracking; Aggregate SMI and Joint BCRB use weighted minimum mean-square error (WMMSE) / minorization–maximization (MM) updates. For Figs.~\ref{fig:traj_close}--\ref{fig:sepsweep}, all designs start from feasible regularized zero-forcing (RZF) communication beams and prior-mean sensing beams. For Figs.~\ref{fig:traj_close} and \ref{fig:traj_well}, F-LMI and the TI designs use a 3000-iteration cap and $10^{-8}$ relative objective-change tolerance, while Aggregate SMI and Joint BCRB use 1000/$10^{-7}$. The separation sweep uses 3000/$10^{-8}$ for all included methods.
% provides a uniform first-order numerical treatment
% The baseline objectives are smooth but do not lead to the closed-form FP block updates derived for F-LMI in Section~\ref{sec:optimization}; PGA therefore provides a uniform first-order numerical treatment under the same power constraint.

\begin{itemize}
\item \emph{Aggregate SMI} \cite{10411942,10255745,10554642,10977963}: \eqref{eq:smi_form} is adapted to the retained sensing outputs by replacing $\mathbf R_x$ with $\mathbf Q_s=\mathbf W_s\mathbf W_s^{\sf H}$ and $L/\sigma_z^2$ with $1/\bar\sigma_{\mathrm{rc}}^2$, with the centered response covariance approximated by 200 common prior samples. It quantifies information about the aggregate response $\sum_{m}\mathbf A(\theta_m)$ and provides no intrinsic stream-to-target association.

\item \emph{TI-SMI}: we construct a target-indexed nominal-response criterion by associating sensing stream $m$ with target $m$ and summing the resulting per-target terms:
\begin{align}
\sum_{m=1}^{M}\log_2\!\left(1+\frac{|\alpha_m|^2}{\bar\sigma_{\mathrm{rc}}^2}\big|\mathbf a^{\sf H}(\mu_{\theta_m})\mathbf w_{s,m}\big|^2\right),
\label{eq:ti_smi}
\end{align}
This baseline uses the same range-compressed normalization as F-LMI, is driven by the nominal response gain at the prior mean, and contains neither the non-intended-target covariance \(\mathbf R_{\mathrm{in},m}\) nor an off-target leakage penalty.

\item \emph{Joint BCRB (D-opt)} \cite{VanTrees2002,10584278}: as the estimation-centered counterpart, the joint Bayesian FIM of all target angles is obtained as 
\begin{align}
\mathbf J_B=\mathbb E_{\boldsymbol\theta}\big[\mathbf F(\boldsymbol\theta)\big]+\boldsymbol\Sigma_\theta^{-1},
\label{eq:joint_bfim}
\end{align}
where $\mathbf F(\boldsymbol\theta)$ is the known-reflectivity angle FIM on the retained sensing outputs, using $\mathbf Q_s$ and $\bar\sigma_{\mathrm{rc}}^2$, and the prior expectation uses 40 common samples. The corresponding joint BCRB is $\mathbf J_B^{-1}$. Since $\mathbf J_B$ is coupled across targets, it must be scalarized before use as a design objective. We adopt the D-optimal criterion $\log\det\mathbf J_B$, equivalently $-\log\det(\mathbf J_B^{-1})$, as a standard scalarization of the joint Bayesian bound; the prior term $\boldsymbol\Sigma_\theta^{-1}$ keeps $\mathbf J_B$ positive definite.

\item \emph{TI-BCRB (A-opt)}: for the noise-only local Gaussian
model of each assigned output, the Bayesian information is
$J_m=\sigma_{\theta_m}^{-2}
+\frac{2}{\bar\sigma_{\mathrm{rc}}^2}
\|\dot{\mathbf A}(\mu_{\theta_m})\mathbf w_{s,m}\|_2^2$,
with corresponding local bound $\mathrm{BCRB}_m=J_m^{-1}$.
We use the prior-normalized A-optimal utility
\begin{align}
\sum_{m=1}^{M}
\Big(1-\frac{\mathrm{BCRB}_m}{\sigma_{\theta_m}^2}\Big),
\label{eq:ti_bcrb}
\end{align}
which sums relative variance reductions and remains bounded.

\item \emph{Joint local MI} (Fig.~\ref{fig:common_basis} only): let
$\mathbf G$ collect the real sensitivities of all retained sensing
outputs to $\Delta\boldsymbol\theta$ at $\boldsymbol\mu_\theta$, and let
$\mathbf C_n$ denote their noise covariance under $\bar\sigma_{\mathrm{rc}}^2$.
The utility is
\begin{align}
\frac12\log_2\det\!\Big(\mathbf I_M+\boldsymbol\Sigma_\theta^{1/2}
\mathbf G^{\sf T}\mathbf C_n^{-1}\mathbf G\,\boldsymbol\Sigma_\theta^{1/2}\Big),
\label{eq:joint_local_mi}
\end{align}
This criterion uses the FIM evaluated at the prior mean instead of the
prior-averaged FIM in \eqref{eq:joint_bfim}.
\end{itemize}

% The two target-indexed baselines isolate distinct design choices: TI-SMI replaces the local angular sensitivity of F-LMI by the nominal response gain, while TI-BCRB retains that sensitivity but measures angular uncertainty through a bounded normalized BCRB reduction rather than a logarithmic information gain. Neither baseline models the prior-induced variation of the non-intended targets.

TI-SMI uses the nominal response gain at the prior mean rather than derivative-based sensitivity and does not explicitly account for the angular prior variance. TI-BCRB retains the derivative sensitivity but uses a bounded BCRB-reduction objective with diminishing marginal value. Neither models the non-intended-target covariance \(\mathbf R_{\mathrm{in},m}\).

The receiver forms the range-compressed statistics in \eqref{eq:rc_model_multi} using the realized communication and sensing sequences. The design objectives rely on local linearization or prior sampling, whereas the evaluation observations are generated from the nonlinear array response with realized symbol blocks and retain every target echo; the operational RMSE therefore evaluates the beamformers beyond their design models.

All physical angles are drawn from untruncated Gaussian priors; out-of-window trials remain in the reported RMSE. Figures~\ref{fig:traj_close} and \ref{fig:traj_well} use nonlinear assigned-output posterior means with multidimensional nuisance integration over the numerical windows in Table~\ref{tab:sim_params}. Figs.~\ref{fig:sepsweep} and \ref{fig:common_basis} use one-dimensional Gaussian-nuisance maximum-likelihood (ML) and posterior-mean receivers, respectively, as detailed in Sections~\ref{subsec:sepsweep} and \ref{subsec:common_basis}. Within each experiment, all methods share channel, angle, and noise realizations. Target-indexed designs retain their stream labels. In Figs.~\ref{fig:traj_close}--\ref{fig:sepsweep}, Aggregate SMI and Joint BCRB use the assignment maximizing $\sum_m|\mathbf a^{\sf H}(\mu_{\theta_m})\mathbf w_{s,\pi(m)}|^2$; Fig.~\ref{fig:common_basis} replaces this with common basis selection. We report the mean channel-wise angular RMSE,
\begin{align}
\overline{\mathrm{RMSE}}
=\frac{1}{N_{\mathrm{ch}}}\sum_{c=1}^{N_{\mathrm{ch}}}
\Bigg(\frac{1}{MN_{\mathrm{tr}}}\sum_{t=1}^{N_{\mathrm{tr}}}\sum_{m=1}^{M}
\big(\hat\theta_{m,c,t}-\theta_{m,c,t}\big)^2\Bigg)^{1/2},
\label{eq:operational_rmse}
\end{align}
where $N_{\mathrm{ch}}$ and $N_{\mathrm{tr}}$ denote the numbers of channel realizations and trials per channel, and angular errors are in degrees; the communication metric is the downlink sum-rate. Design uses $\bar\sigma_{\mathrm{rc}}^2=\sigma_z^2/(L-K-M)$, while inference uses the realized conditional LS-noise covariance. At matched-rate comparisons, designs are time-shared per channel, mixing squared errors inside \eqref{eq:operational_rmse} before taking the channel-wise root, without power rescaling. For Figs.~\ref{fig:traj_close} and \ref{fig:traj_well}, the matched-rate summaries use three-fold cross-fitting over common sensing trials. In each fold, among all feasible single designs and two-design mixtures of each method's nine saved beamformers, the candidate with the lowest MSE on the other two folds is selected and evaluated on the held-out fold. The held-out squared errors from all folds are combined before taking the channel-wise root. Per-channel rate targets are fixed to the reference F-LMI designs, without clipping or channel removal. This is a validation-based selection over saved designs rather than a globally optimal frontier. At the lower rate of each layout, only TI-BCRB among the baselines has saved operating points bracketing the target rate in every channel, so only that comparison is reported.

For F-LMI, the block tradeoff parameter $\lambda$ is implemented by setting $a_{c,k}=1-\lambda$ and $b_{s,m}=L\lambda$ for all $k$ and $m$, giving $(1-\lambda)R_C+\lambda\sum_m I_m$, where $R_C\triangleq\sum_k R_k$. Figures~\ref{fig:traj_close} and \ref{fig:traj_well} use nine weights per method. The target-separation sweep fixes each channel's communication-rate target to the saved $s=3^\circ$ F-LMI design at $\lambda=0.9$.

\subsection{Operational Rate--RMSE Trajectories} 
\begin{figure}[!t]
\centering
\includegraphics[width=0.95\columnwidth]{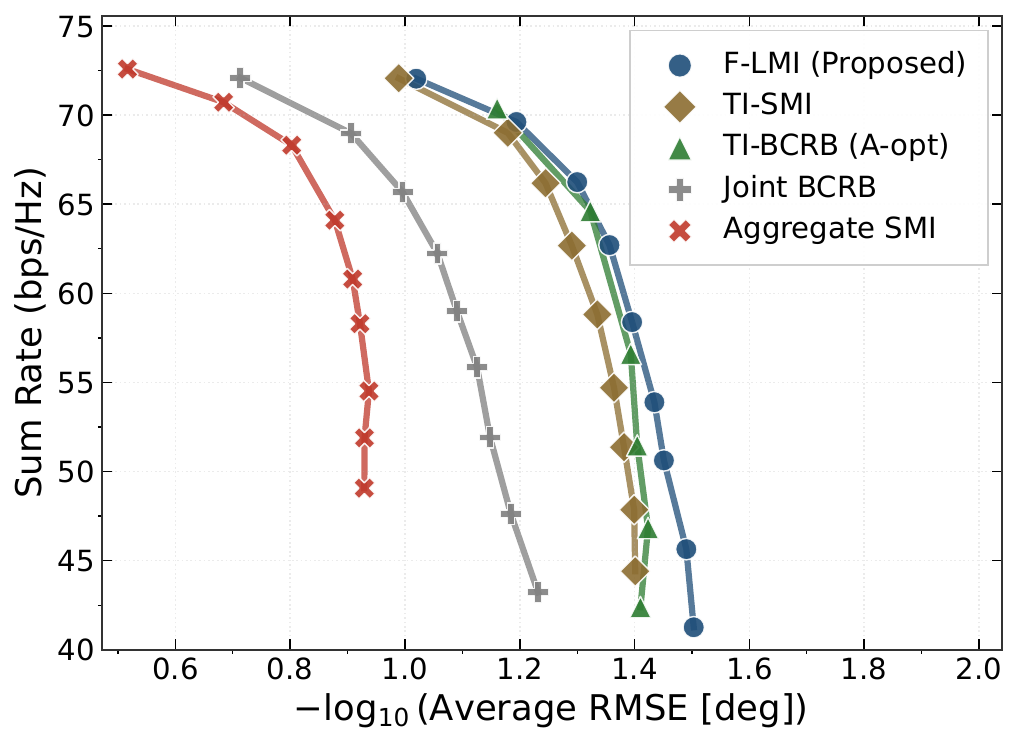}
\caption{Operational rate--RMSE trajectories for the closely spaced target
layout $\{-3^\circ,0^\circ,3^\circ\}$, $\sigma_\theta=0.5^\circ$, with nonlinear assigned-output posterior-mean estimation. Larger horizontal values indicate lower average angular RMSE; only rates above 40~bps/Hz are displayed.}
\label{fig:traj_close}
\end{figure}

\begin{figure}[!t]
\centering
\includegraphics[width=0.95\columnwidth]{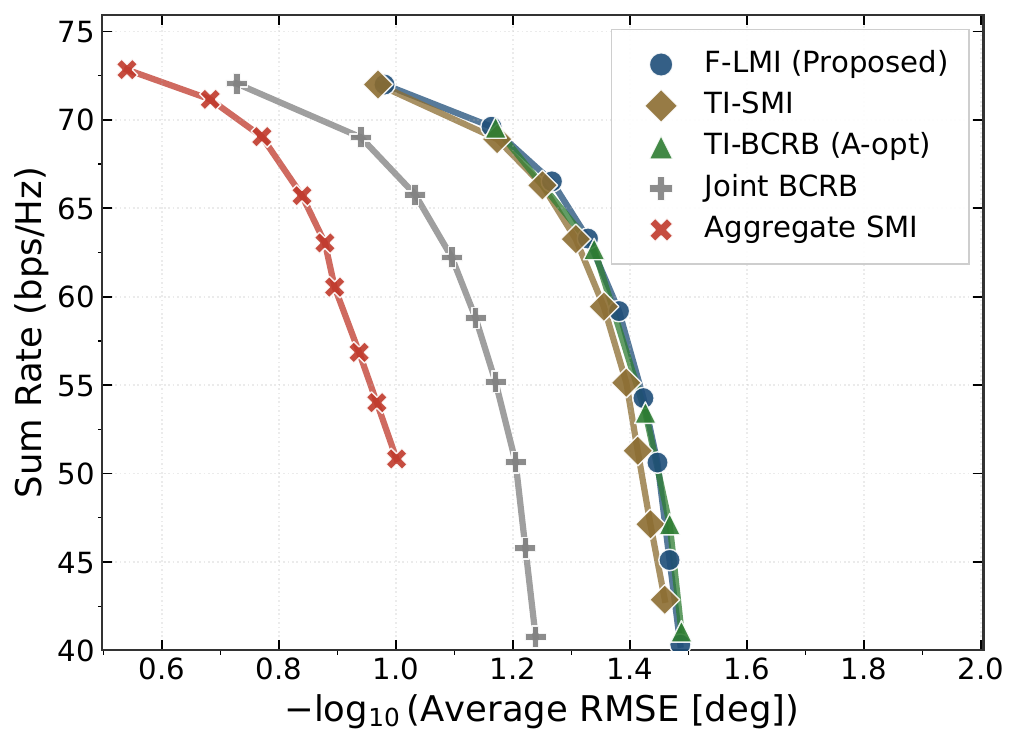}
\caption{Operational rate--RMSE trajectories for the well-separated target layout
$\{-40^\circ,0^\circ,40^\circ\}$, with the same $0.5^\circ$ prior and nonlinear assigned-output posterior-mean receiver as Fig.~\ref{fig:traj_close}.}
\label{fig:traj_well}
\end{figure}

Markers in Figs.~\ref{fig:traj_close} and \ref{fig:traj_well} show individual weight-swept designs; the matched-rate comparisons below use the cross-fitted selection and time sharing described in Section~\ref{subsec:evaluation_protocol}.
% The curves in Figs.~\ref{fig:traj_close}--\ref{fig:traj_well} connect
% weight-swept operating points and should be interpreted as trajectories
% rather than Pareto frontiers.

Fig.~\ref{fig:traj_close} shows the closely spaced target case. F-LMI improves the rate--RMSE tradeoff over the target-indexed baselines at representative matched rates: in paired comparisons, it reduces the RMSE by $14.7\%$ and $5.0\%$ relative to TI-SMI and TI-BCRB at $62.71$~bps/Hz, and by $14.7\%$ relative to TI-BCRB at $45.64$~bps/Hz. The paired 95\% confidence intervals (CIs) of all three differences exclude zero.

The target-indexed baselines improve substantially over Aggregate SMI and Joint BCRB under this assigned-output receiver. Aggregate SMI levels off, whereas Joint BCRB continues to improve as more rate is sacrificed for sensing. The remaining gap to F-LMI is consistent with the value of combining stream-to-target association, derivative sensitivity, and prior-induced nuisance covariance; the last component matters when the receive-sensitivity subspaces permit stronger overlap, as at $3^\circ$ spacing. The covariance-based designs' errors here are conditional on their stream factorization and assignment, which Section~\ref{subsec:common_basis} examines.

% In contrast, Aggregate SMI and Joint BCRB can sacrifice communication rate without producing a proportional improvement in the operational RMSE, reflecting a mismatch between aggregate or joint sensing objectives and the stream-wise matched-filtered statistics used by the receiver.

Fig.~\ref{fig:traj_well} reports the well-separated target case. At a matched mean rate of $63.29$~bps/Hz, F-LMI reduces the RMSE by $5.7\%$ and $3.6\%$ relative to TI-SMI and TI-BCRB, with both paired 95\% CIs excluding zero. At $45.11$~bps/Hz, the paired 95\% CI of the F-LMI--TI-BCRB difference includes zero.

Both F-LMI and TI-BCRB use intended angular derivatives. For any beamformer, $\mathbf J_{\ell,m}$ lies in the complex span of $\mathbf b(\mu_{\theta_\ell})$ and $\dot{\mathbf b}(\mu_{\theta_\ell})$. For the 64-element receive ULA, the largest principal-angle cosine between the real-stacked spans of the intended and neighboring targets, maximized over target pairs, therefore bounds the alignment of $\tilde{\mathbf J}_{m,m}$ and $\tilde{\mathbf J}_{\ell,m}$; it is about $0.05$ at $40^\circ$ spacing, compared with $0.43$ at $3^\circ$. The noise-only bound on the quadratic form in \eqref{eq:flmi_marginal_compact} is nearly attained, and F-LMI differs from TI-BCRB mainly through the scalarization of the same per-target information. Joint BCRB and Aggregate SMI attain higher assigned-output RMSE than the target-indexed designs. The F-LMI advantage over TI-BCRB is therefore larger when neighboring responses overlap.

\subsection{Sensitivity to Target Separation}
\label{subsec:sepsweep}

\begin{figure}[!t]
\centering
\includegraphics[width=0.95\columnwidth] {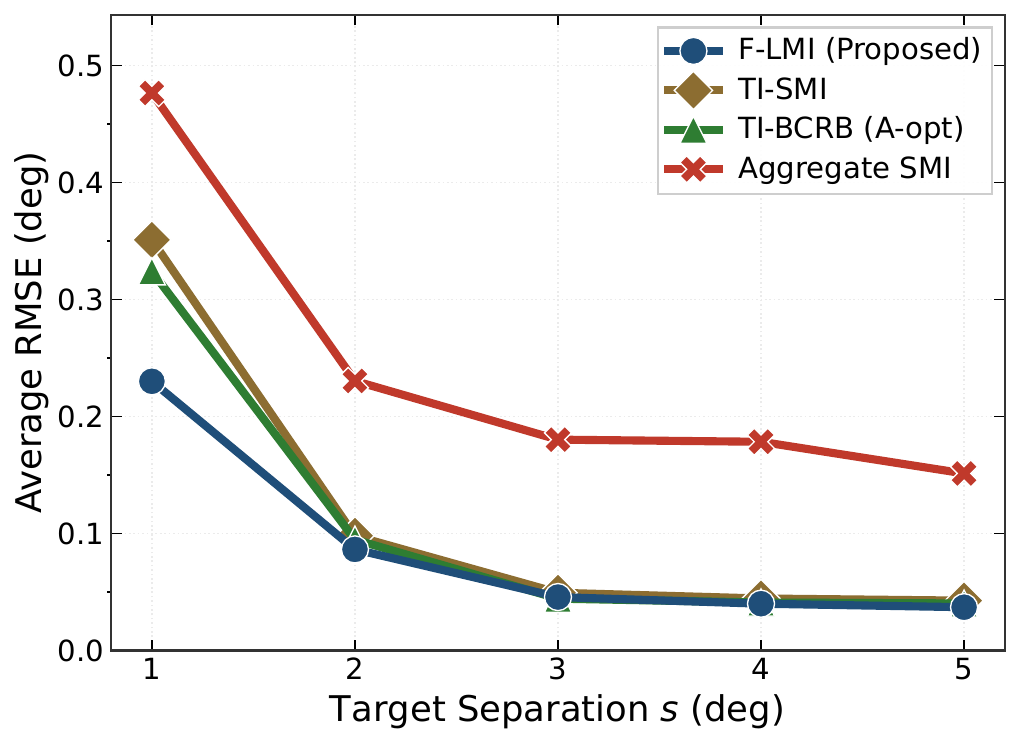}
\caption{Average angular RMSE versus target separation for $\{-s,0,s\}^\circ$, $s\in\{1,2,3,4,5\}$, with $\sigma_\theta=0.5^\circ$. All designs use the same branch ML receiver and matched per-channel communication rates, without power rescaling.}
\label{fig:sepsweep}
\end{figure}

To quantify the effect of target geometry, Fig.~\ref{fig:sepsweep} uses $s\in\{1,2,3,4,5\}^\circ$ in the layout $\{-s,0,s\}^\circ$. All separations share the same communication-channel realizations, prior-deviation and LS-noise draws, priors, power budget, and initialization rule. Each channel's rate target is fixed to its $s=3^\circ$ F-LMI design at $\lambda=0.9$ (mean $56.39$~bps/Hz) and matched by feasible time sharing without sensing-power normalization.

All methods use the same branch-wise one-dimensional Gaussian-surrogate ML receiver: each branch retains the nonlinear intended response, models the other targets through their prior-induced mean and covariance (64-point Gaussian quadrature), and searches $[-1.50^\circ,1.47^\circ]$ around the nominal angle in $0.03^\circ$ steps without an intended-angle prior penalty. Confidence intervals are obtained from 4000 paired channel-bootstrap resamples.

All four designs show decreasing mean RMSE over the tested range. F-LMI reduces RMSE relative to TI-BCRB by $29.0\%$, $8.8\%$, and $8.0\%$ at $1^\circ$, $2^\circ$, and $5^\circ$, respectively, with paired 95\% difference CIs excluding zero. At $3^\circ$, TI-BCRB has a slightly lower mean; the difference CIs include zero at both $3^\circ$ and $4^\circ$. The alignment bounds of $0.998$ at $1^\circ$ and $0.88$ at $2^\circ$ permit strong nuisance coupling. Increasing separation also changes angular sensitivity and power allocation, so the relative gain need not decrease monotonically. The comparison reflects both $\mathbf R_{\mathrm{in},m}$ and logarithmic scalarization.

\subsection{Comparison with Common Basis Selection}
\label{subsec:common_basis}

\begin{figure}[!t]
\centering
\includegraphics[width=0.95\columnwidth]{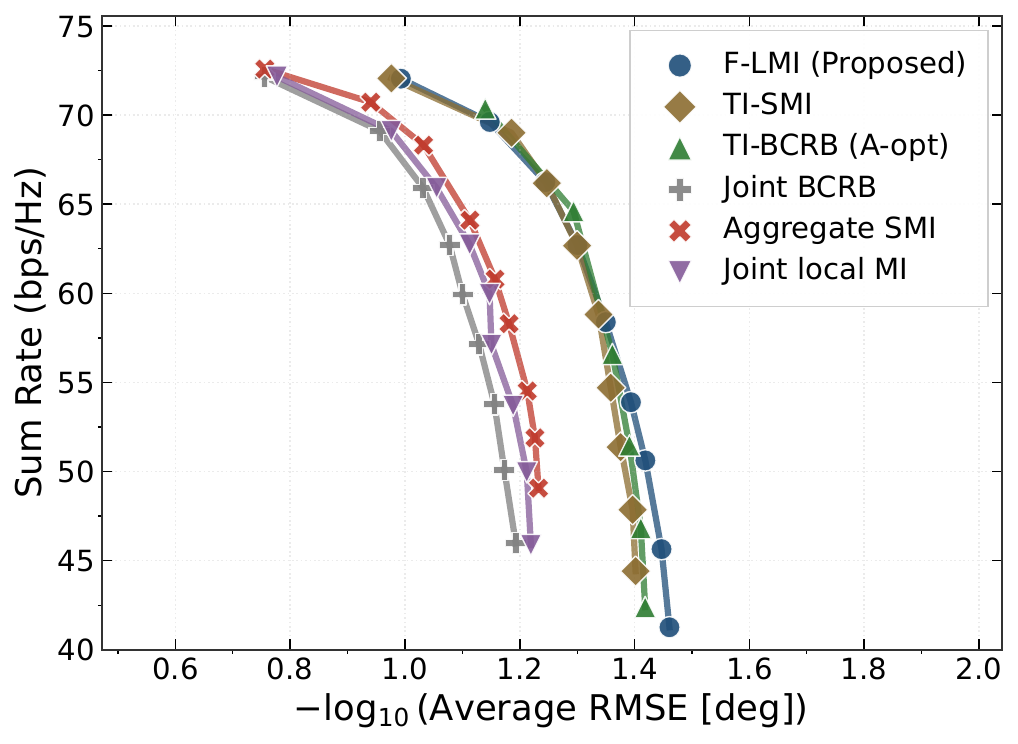}
\caption{Rate--RMSE trajectories for the closely spaced layout under a
common basis-selection protocol, using a Gaussian-nuisance branch
posterior-mean receiver. Basis selection and final evaluation use
separate realizations.}
\label{fig:common_basis}
\end{figure}

Aggregate SMI and Joint BCRB depend on $\mathbf W_s$ only through
$\mathbf W_s\mathbf W_s^{\sf H}$ and are therefore invariant to
$\mathbf W_s\mapsto\mathbf W_s\mathbf U$ for unitary $\mathbf U$, whereas the
assigned-output receiver depends on the individual beam columns.
Fig.~\ref{fig:common_basis} therefore applies a common basis-selection
protocol to every design and adds the Joint local MI baseline in
\eqref{eq:joint_local_mi}. Joint local MI shares the prior-mean
linearization with F-LMI but evaluates all retained sensing outputs
jointly instead of assigning a marginal term to each target's branch, and
thus serves as a local-model reference for the branch-wise objective.
Since every design is evaluated with the same branch receiver, this
experiment compares beamforming criteria rather than joint versus
branch-wise estimation.
For each channel, design, and operating point, we evaluate
107 bases: six permutations (including the original basis),
100 Haar-random unitary bases with derivative-based assignment,
and one basis from four-start prior-based F-LMI optimization.
This optimization varies only $\mathbf U$, keeping
$\mathbf W_c$ and $\mathbf Q_s$ fixed. Selection minimizes branch MSE on independent validation realizations simulated from the assumed angular priors and observation model. These realizations are separate from those used for final evaluation. All methods use the same candidate construction and selection rule.

Joint BCRB and Joint local MI are reoptimized with four
starts, a 6000-iteration cap, and a $10^{-11}$ relative objective-change
tolerance, keeping the
best start by the design objective; the other methods retain their saved
beamformers. The receiver is a one-dimensional branch posterior-mean
estimator using the nonlinear intended response and Gaussian nuisance
moments (64-point Gauss--Hermite quadrature), with an intended-angle grid
$\mu_{\theta_m}\pm6\sigma_{\theta_m}$ in $0.03^\circ$ steps.

At a matched mean rate of $59.86$~bps/Hz, F-LMI reduces branch RMSE by $31.8\%$, $35.7\%$, and $41.6\%$ relative to Aggregate SMI, Joint local MI, and Joint BCRB, respectively; the paired 95\% CI of its difference from TI-BCRB includes zero. These results use the stated finite basis search; differences from Fig.~\ref{fig:traj_close} also reflect the receiver and reoptimized joint designs. Basis selection redistributes target responses across branches
while preserving covariance-based objectives. The remaining gaps
under the tested basis search support matching the utility to
the information available in each assigned output.

\section{Conclusion}
\label{sec:conclusion}

% This paper proposed F-LMI, a prior-aware local MI utility for
% MU-MIMO ISAC beamforming. By linearizing target-specific
% matched-filtered sensing statistics around the angular prior
% mean, F-LMI yields a closed-form bit-domain metric that is
% aligned with local Bayesian estimation accuracy. The
% multi-target extension uses a receiver-oriented separable
% construction in which non-intended target variations are
% modeled as structured interference. We further established an explicit non-asymptotic bound relating F-LMI to the nonlinear parameter MI and to the BCRB log-ratio in a non-degenerate local regime.

We developed F-LMI for MU-MIMO ISAC beamforming by combining
angular sensitivity with prior-induced nuisance covariance.
Fixed-beamformer single-target analysis establishes local
connections to nonlinear MI and a BCRB log-ratio, while FP
provides linear-system beam updates. The separable utility also supports target-specific sensing priorities through its per-target weights. Monte Carlo simulations with nonlinear sensing observations show rate-matched gains over TI-BCRB for closely spaced targets and lower branch RMSE than aggregate and joint criteria under common basis selection. The separation sweep shows the largest relative gain over TI-BCRB at the smallest tested target separation. The common-basis comparison further supports matching the sensing utility to the information retained by each assigned output. Extensions include self-interference-aware beamforming \cite{BayraktarTWC26_FDISAC}, joint nonlinear inference, uncertain reflectivities, and clutter priors.
%XXXX

\bibliographystyle{IEEEtran}
\bibliography{FLMI_ref}

@ARTICLE{10584278,
    author={Xu, Chan and Zhang, Shuowen},
    journal={IEEE J. Sel. Areas Commun.}, 
    title={{MIMO} Integrated Sensing and Communication Exploiting Prior Information}, 
    year={2024},
    volume={42},
    number={9},
    pages={2306-2321},
    doi={10.1109/JSAC.2024.3413972}
}

@ARTICLE{10411942,
    author={Li, Jin and Zhou, Gui and Gong, Tantao and Liu, Nan},
    journal={IEEE Trans. Veh. Technol.}, 
    title={A Framework for Mutual Information-Based {MIMO} Integrated Sensing and Communication Beamforming Design}, 
    year={2024},
    volume={73},
    number={6},
    pages={8352-8366},
    doi={10.1109/TVT.2024.3355899}
}

@ARTICLE{1412024,
    author={Dongning Guo and Shamai, S. and Verdu, S.},
    journal={IEEE Trans. Inf. Theory}, 
    title={Mutual information and minimum mean-square error in {Gaussian} channels}, 
    year={2005},
    volume={51},
    number={4},
    pages={1261-1282},
    doi={10.1109/TIT.2005.844072}
}

@ARTICLE{10554642,
    author={Peng, Yizhou and Yang, Songjie and Lyu, Wanting and Li, Ya and He, Hongjun and Zhang, Zhongpei and Assi, Chadi},
    journal={IEEE Wireless Commun. Lett.}, 
    title={Mutual Information-Based Integrated Sensing and Communications: A {WMMSE} Framework}, 
    year={2024},
    volume={13},
    number={10},
    pages={2642-2646},
    doi={10.1109/LWC.2024.3412974}
}

@ARTICLE{10255745,
    author={Wei, Zhiqing and Piao, Jinghui and Yuan, Xin and Wu, Huici and Zhang, J. Andrew and Feng, Zhiyong and Wang, Lin and Zhang, Ping},
    journal={IEEE Trans. Commun.}, 
    title={Waveform Design for {MIMO}-{OFDM} Integrated Sensing and Communication System: An Information Theoretical Approach}, 
    year={2024},
    volume={72},
    number={1},
    pages={496-509},
    doi={10.1109/TCOMM.2023.3317258}
}

@article{LiuTWC18,
    author={Liu, Fan and Masouros, Christos and Li, Ang and Sun, Huafei and Hanzo, Lajos},
    journal={IEEE Trans. Wireless Commun.}, 
    title={{MU-MIMO} communications with {MIMO} Radar: From Co-Existence to Joint Transmission}, 
    year={2018},
    volume={17},
    number={4},
    pages={2755-2770},
    month= Apr,
}

@article{LiuJSAC2022_ISAC,
    author  = {Fan Liu and Yuanhao Cui and Christos Masouros and Jie Xu and Tony Xiao Han and Yonina C. Eldar and Stefano Buzzi},
    title   = {Integrated Sensing and Communications: Towards Dual-functional Wireless Networks for {6G} and Beyond},
    journal = {IEEE J. Sel. Areas Commun.},
    volume  = {40},
    number  = {6},
    pages   = {1728--1767},
    month   = jun,
    year    = {2022},
    doi     = {10.1109/JSAC.2022.3156632}
}

@book{Schreier2010,
    author    = {Peter J. Schreier and Louis L. Scharf},
    title     = {Statistical Signal Processing of Complex-Valued Data: The Theory of Improper and Noncircular Complex Signals},
    publisher = {Cambridge University Press},
    year      = {2010},
    address   = {Cambridge, UK},
    doi       = {10.1017/CBO9780511815911}
}

@article{HuaTVT2023_TypeII,
    author  = {Haocheng Hua and Jie Xu and Tony Xiao Han},
    title   = {Optimal Transmit Beamforming for Integrated Sensing and Communication},
    journal = {IEEE Trans. Veh. Technol.},
    volume  = {72},
    number  = {8},
    pages   = {10588--10603},
    year    = {2023},
    doi     = {10.1109/TVT.2023.3262513}
}

@article{liu_tsp2020_joint,
    author  = {Xiang Liu and Tianyao Huang and Nir Shlezinger and Yimin Liu and Jie Zhou and Yonina C. Eldar},
    title   = {Joint Transmit Beamforming for Multiuser {MIMO} Communications and {MIMO} Radar},
    journal = {IEEE Trans. Signal Process.},
    volume  = {68},
    pages   = {3929--3944},
    year    = {2020},
    doi     = {10.1109/TSP.2020.3004739}
}

@article{liu_tcomm2020_road,
    author  = {Fan Liu and Christos Masouros and Athina P. Petropulu and Hugh Griffiths and Lajos Hanzo},
    title   = {Joint Radar and Communication Design: Applications, State-of-the-Art, and the Road Ahead},
    journal = {IEEE Trans. Commun.},
    volume  = {68},
    number  = {6},
    pages   = {3834--3862},
    year    = {2020},
    month   = jun,
    doi     = {10.1109/TCOMM.2020.2973976}
}

@article{xiong_tit2023_gaussian,
    author  = {Yifeng Xiong and Fan Liu and Yuanhao Cui and Weijie Yuan and Tony X. Han and Giuseppe Caire},
    title   = {On the Fundamental Tradeoff of Integrated Sensing and Communications Under {Gaussian} Channels},
    journal = {IEEE Trans. Inf. Theory},
    volume  = {69},
    number  = {9},
    pages   = {5723--5751},
    year    = {2023},
    month   = sep,
    doi     = {10.1109/TIT.2023.3284449}
}

@article{LiuTSP22_CRB_JRC,
    author  = {Fan Liu and Ya-Feng Liu and Ang Li and Christos Masouros and Yonina C. Eldar},
    title   = {{Cram\'er--Rao} bound optimization for Joint Radar-Communication Beamforming},
    journal = {IEEE Trans. Signal Process.},
    volume  = {70},
    pages   = {240--253},
    year    = {2022},
    doi     = {10.1109/TSP.2021.3135692}
}

@article{HuaTWC24_CRBRate,
    author  = {Haocheng Hua and Tony Xiao Han and Jie Xu},
    title   = {{MIMO} Integrated Sensing and Communication: {CRB}-Rate Tradeoff},
    journal = {IEEE Trans. Wireless Commun.},
    volume  = {23},
    number  = {4},
    pages   = {2839--2854},
    year    = {2024},
    month   = apr,
    doi     = {10.1109/TWC.2023.3303326}
}

@article{RenTWC24_MultiTargetCRBRate,
    author  = {Zixiang Ren and Yunfei Peng and Xianxin Song and Yuan Fang and Ling Qiu and Liang Liu and Derrick Wing Kwan Ng and Jie Xu},
    title   = {Fundamental {CRB}-Rate Tradeoff in Multi-Antenna {ISAC} Systems with Information Multicasting and Multi-Target Sensing},
    journal = {IEEE Trans. Wireless Commun.},
    volume  = {23},
    number  = {4},
    pages   = {3870--3885},
    year    = {2024},
    month   = apr,
    doi     = {10.1109/TWC.2023.3312723}
}

@article{ChenTWC25_FastFP_FI,
    author  = {Yannan Chen and Yi Feng and Xiaoyang Li and Licheng Zhao and Kaiming Shen},
    title   = {Fast Fractional Programming for Multi-Cell Integrated Sensing and Communications},
    journal = {IEEE Trans. Wireless Commun.},
    volume  = {24},
    number  = {8},
    pages   = {6797--6812},
    year    = {2025},
    doi     = {10.1109/TWC.2025.3556301}
}

@article{LiTSP08_CRB_MIMORadar,
    author  = {Jian Li and Luzhou Xu and Petre Stoica and K. W. Forsythe and David W. Bliss},
    title = {Range compression and waveform optimization for {MIMO} radar: A {Cram\'er--Rao} bound based study},
    journal = {IEEE Trans. Signal Process.},
    volume  = {56},
    number  = {1},
    pages   = {218--232},
    year    = {2008},
    month   = jan,
    doi     = {10.1109/TSP.2007.901653}
}

@article{YangBlumTAES07_MI_MMSE,
    author  = {Y. Yang and R. S. Blum},
    title   = {{MIMO} Radar Waveform Design Based on Mutual Information and Minimum Mean-Square Error Estimation},
    journal = {IEEE Trans. Aerosp. Electron. Syst.},
    volume  = {43},
    number  = {1},
    pages   = {330--343},
    year    = {2007},
    doi     = {10.1109/TAES.2007.357137}
}

@article{ShenTWC18,
  author  = {K. Shen and W. Yu},
  title   = {Fractional programming for communication systems---{Part I}: Power control and beamforming},
  journal = {IEEE Trans. Signal Process.},
  volume  = {66},
  number  = {10},
  pages   = {2616--2630},
  month   = {May},
  year    = {2018}
}

@article{LiuTSP18_DFRC_OptimalWaveform,
    author  = {Fan Liu and Longfei Zhou and Christos Masouros and Ang Li and Wu Luo and Athina Petropulu},
    title   = {Toward Dual-Functional Radar-Communication Systems: Optimal Waveform Design},
    journal = {IEEE Trans. Signal Process.},
    volume  = {66},
    number  = {16},
    pages   = {4264--4279},
    year    = {2018},
    doi     = {10.1109/TSP.2018.2847648}
}

@book{VanTrees2002,
    author = {{H. L. Van Trees}},
    title     = {Detection, Estimation, and Modulation Theory, Part IV: Optimum Array Processing},
    publisher = {John Wiley \& Sons},
    year      = {2002},
    address   = {New York, NY, USA}
}

@article{ShenToN19_MatrixFP,
    author  = {Kaiming Shen and Wei Yu and Licheng Zhao and Daniel P. Palomar},
    title   = {Optimization of {{MIMO}} Device-to-Device Networks via Matrix Fractional Programming: A Minorization-Maximization Approach},
    journal = {IEEE/ACM Trans. Netw.},
    volume  = {27},
    number  = {5},
    pages   = {2164--2177},
    year    = {2019},
    doi     = {10.1109/TNET.2019.2935352}
}

@ARTICLE{4350230,
    author={Li, Jian and Stoica, Petre},
    journal={IEEE Signal Process. Mag.}, 
    title={{MIMO} Radar with Colocated Antennas}, 
    year={2007},
    volume={24},
    number={5},
    pages={106-114},
    doi={10.1109/MSP.2007.904812}
}

@ARTICLE{6541985,
    author={Huleihel, Wasim and Tabrikian, Joseph and Shavit, Reuven},
    journal={IEEE Trans. Signal Process.}, 
    title={Optimal Adaptive Waveform Design for Cognitive {MIMO} Radar}, 
    year={2013},
    volume={61},
    number={20},
    pages={5075-5089},
    doi={10.1109/TSP.2013.2269045}
}

@ARTICLE{4276989,
    author={Stoica, Petre and Li, Jian and Xie, Yao},
    journal={IEEE Trans. Signal Process.}, 
    title={On Probing Signal Design For {MIMO} Radar}, 
    year={2007},
    volume={55},
    number={8},
    pages={4151-4161},
    doi={10.1109/TSP.2007.894398}
}

@ARTICLE{1564431,
    author={Palomar, D.P. and Verdu, S.},
    journal={IEEE Trans. Inf. Theory}, 
    title={Gradient of mutual information in linear vector {Gaussian} channels}, 
    year={2006},
    volume={52},
    number={1},
    pages={141-154},
    doi={10.1109/TIT.2005.860424}
}

@article{ShiuTCOM2000_OneRing,
  author  = {D. Shiu and G. J. Foschini and M. J. Gans and J. M. Kahn},
  title   = {Fading correlation and its effect on the capacity of multielement antenna systems},
  journal = {IEEE Trans. Commun.},
  volume  = {48}, number = {3}, pages = {502--513}, year = {2000}
}

@article{SMIUB,
  author  = {Meng, Chunwei and Wei, Zhiqing and Ma, Dingyou and Ni, Wanli and Su, Liyan and Feng, Zhiyong},
  title   = {Multi-Objective Optimization-based Transmit Beamforming for Multi-Target and Multi-User {MIMO}-{ISAC} Systems},
  journal = {arXiv preprint arXiv:2405.09022},
  year    = {2024}
}

@article{ChiriyathTSP16_EstRate,
  author  = {Chiriyath, Alex R. and Paul, Bryan and Jacyna, Garry M. and Bliss, Daniel W.},
  title   = {Inner bounds on performance of radar and communications co-existence},
  journal = {IEEE Trans. Signal Process.},
  volume  = {64},
  number  = {2},
  pages   = {464--474},
  month   = jan,
  year    = {2016}
}

@article{ChiriyathTCCN17_Convergence,
  author  = {Chiriyath, Alex R. and Paul, Bryan and Bliss, Daniel W.},
  title   = {Radar-communications convergence: {C}oexistence, cooperation, and co-design},
  journal = {IEEE Trans. Cogn. Commun. Netw.},
  volume  = {3},
  number  = {1},
  pages   = {1--12},
  month   = mar,
  year    = {2017}
}

@article{DongTVT23_EstRate,
  author  = {Dong, Fuwang and Liu, Fan and Lu, Shihang and Xiong, Yifeng},
  title   = {Rethinking estimation rate for wireless sensing: {A} rate-distortion perspective},
  journal = {IEEE Trans. Veh. Technol.},
  volume  = {72},
  number  = {12},
  pages   = {16876--16881},
  month   = dec,
  year    = {2023}
}

@article{ChalonerVerdinelli95,
  author  = {Chaloner, Kathryn and Verdinelli, Isabella},
  title   = {{B}ayesian experimental design: {A} review},
  journal = {Statist. Sci.},
  volume  = {10},
  number  = {3},
  pages   = {273--304},
  year    = {1995}
}

@misc{Alexanderian22_Dopt,
  author       = {Alexanderian, Alen},
  title        = {A brief note on the {Bayesian}
                  {D}-optimality criterion},
  howpublished = {arXiv preprint arXiv:2212.11466},
  year         = {2023}
}

@article{Maiwald2000_InvWishart,
  author  = {Maiwald, Dirk and Kraus, Dieter},
  title   = {Calculation of moments of complex {W}ishart and complex inverse {W}ishart distributed matrices},
  journal = {IEE Proc. Radar, Sonar Navig.},
  volume  = {147},
  number  = {4},
  pages   = {162--168},
  month   = aug,
  year    = {2000}
}

@inproceedings{YaoZhang_MultiTarget,
  author    = {Yao, Jiayi and Zhang, Shuowen},
  title     = {Optimal Transmit Signal Design for Multi-Target
               {MIMO} Sensing Exploiting Prior Information},
  booktitle = {Proc. IEEE GLOBECOM},
  pages     = {4920--4925},
  year      = {2024},
  doi       = {10.1109/GLOBECOM52923.2024.10901183}
}

@article{YaoZhang_HowManyBeams,
  author  = {Yao, Jiayi and Zhang, Shuowen},
  title   = {Optimal Beamforming for Integrated Sensing and
             Communication Exploiting Prior Distribution Information:
             How Many Beams are Needed?},
  journal = {arXiv preprint arXiv:2503.03560v4},
  year    = {2026}
}

@ARTICLE{BayraktarTWC26_FDISAC,
  author={Bayraktar, Murat and González-Prelcic, Nuria and Valkama, Mikko and Chen, Hao and Jianzhong Zhang, Charlie},
  journal={IEEE Trans. Wireless Commun.}, 
  title={Integrated Monostatic Sensing and Full-Duplex Multiuser Communication for {mmWave} Systems}, 
  year={2026},
  volume={25},
  number={},
  pages={10904-10920},
  doi={10.1109/TWC.2026.3653009}}

@ARTICLE{10977963,
  author={Xie, Lei and Liu, Fan and Luo, Jiajin and Song, Shenghui},
  journal={IEEE Trans. Commun.}, 
  title={Sensing Mutual Information With Random Signals in {G}aussian Channels}, 
  year={2025},
  volume={73},
  number={10},
  pages={9437-9452},
  doi={10.1109/TCOMM.2025.3564767}}
\end{document}